\documentclass[prd,aps,onecolumn,showpacs,nofootinbib,preprintnumbers]{revtex4}

\usepackage{amsmath,amssymb,graphicx}
\usepackage{graphicx}
\usepackage{epsfig}
\usepackage{float}
\usepackage[usenames]{color}

\newcommand{\be}{\begin{equation}}
\newcommand{\ee}{\end{equation}}
\newcommand{\bea}{\begin{eqnarray}}
\newcommand{\eea}{\end{eqnarray}}
\def\ba#1\ea{\begin{align}#1\end{align}}

\newcommand{\vs}{\nonumber\\}

\def\bfu{\mathbf{u}}
\def\bfk{\mathbf{k}}
\def\bfq{\mathbf{q}}

\definecolor{RedWine}{rgb}{0.743,0,0}

\begin{document}

\title{Non-linear power spectra in the synchronous gauge}
\author{Jai-chan Hwang${}^{1}$, Hyerim Noh${}^{2}$, Donghui Jeong${}^{3,4}$, Jinn-Ouk Gong${}^{5,6}$, and Sang Gyu Biern${}^{7}$}
\affiliation{${}^{1}$Department of Astronomy and Atmospheric Sciences, Kyungpook National University, Daegu, 702-701, Korea \\
${}^{2}$Korea Astronomy and Space Science Institute, Daejeon, 305-348, Korea \\
${}^{3}$Department of Astronomy and Astrophysics,
The Pennsylvania State University, University Park, PA 16802, USA\\
${}^{4}$Institute for Gravitation and the Cosmos, The Pennsylvania State University, University Park, PA 16802, USA\\
${}^{5}$ Asia Pacific Center for Theoretical Physics, Pohang 790-784, Korea \\
${}^{6}$ Department of Physics, Postech, Pohang 790-784, Korea \\
${}^{7}$ Department of Physics, Seoul National University, Seoul 151-747, Korea}
\preprint{\vtop{\hbox{APCTP-Pre2014-012}}}

\date{\today}

\begin{abstract}

 We study the non-linear corrections to the matter and velocity 
 power spectra in the synchronous gauge (SG). We consider the perturbations up 
 to third order in a zero-pressure fluid in flat cosmological background, 
 which is relevant for the non-linear growth of cosmic structure. 
 As a result, we point out that the SG is an inappropriate coordinate 
 choice when handling the non-linear growth of the large-scale structure.
 Although the equations in the SG happen to coincide with those in the comoving
 gauge (CG) to linear order, they differ from second order. 
 In particular, the second order hydrodynamic equations in the the SG are 
 apparently in the Lagrangian form, whereas those in the CG are in the 
 Eulerian form. Thus, the non-linear power spectra naively presented in the 
 original SG show strange behavior quite different from the result of 
 the Newtonian theory even on sub-horizon scales.
 The power spectra in the SG show regularized behaviors only after 
 we introduce convective terms in the second order so that the equations
 in two gauges coincide to the second order.

\end{abstract}

\noindent 

\maketitle

\section{Introduction}

The original study of the linearized cosmological perturbation in Einstein's 
gravity was made in the synchronous gauge (SG) \cite{Lifshitz-1946}.
Despite its shortcoming of leaving remnant gauge mode even after imposing the gauge condition, the SG is still popularly used in the literature 
(See, e.g. \cite{Peebles-1980}).
For a zero-pressure medium, the linear order equations in the SG coincide 
with those in the comoving gauge (CG) \cite{Nariai-1969,Bardeen-1980}.
It is because the remnant gauge mode present in the SG happens to coincide 
with one of the physical solutions in the zero-pressure medium.
For the general background medium, however, the remnant gauge mode in the SG 
reveals its own identity (different from the physical solution in the SG). 
The example in the medium with pressure is shown in \cite{Lifshitz-1946}.
For a zero-pressure medium in both gauges, when identifying the perturbed 
expansion scalar as the peculiar velocity, the linear equations for the 
density and velocity perturbations coincide with the Newtonian counterparts.
Therefore, in a zero-pressure medium the linear matter and velocity power spectra in the SG coincide with those in the CG, or the Newtonian theory.

Situation is quite different, however, as we consider non-linear perturbations.
To non-linear order, the relativistic/Newtonian correspondences are known 
for a zero-pressure fluid: the Newtonian limit is available as the 
infinite speed-of-light limit of Einstein's gravity in both the zero-shear 
gauge (often known as the longitudinal, the conformal-Newtonian, or the 
Poisson gauge) and the uniform-expansion gauge 
(often known as the uniform-Hubble gauge) \cite{Hwang-Noh-2013-Newtonian}.
As we can think of the infinite speed-of-light limit as the case when all
modes are subhorizon, the Newtonian limit we have listed above is possible 
only in the subhorizon limit, but is valid to fully non-linear order for 
the density, velocity and the gravitational potential.

In the CG, up to second order the relativistic perturbation 
equations for density and velocity coincide with the Newtonian hydrodynamic 
equations when replacing the perturbed gravitational potential by using 
the Poisson equation \cite{Noh-Hwang-2004}.
This correspondence is valid in all scales.
In handling the density and velocity perturbations, the pure Einstein's gravity correction terms start appearing from third order 
in the CG \cite{Hwang-Noh-2005-third}.
As the next-to-leading order power spectrum of Gaussian field demands 
perturbation to third order \cite{Vishniac-1983}, we expect 
pure Einstein's gravity corrections to the non-linear power spectrum in the CG.
Such corrections, however, in the CG are well regularized and quite 
suppressed on all scales compared to the \textit{Newtonian} terms \cite{JGNH-2011}.

Now, in the SG, the hydrodynamic equations show similarity with the Newtonian ones in the Lagrangian frame whereas those in the CG show exact correspondence with the Newtonian equations in the Eulerian frame, and such a difference appears from the second order in perturbation \cite{Hwang-Noh-2006-SG}, see Section~\ref{sec:fully-NL}.
In this work our aim is to present the next-to-leading order density and velocity power spectra in the SG.
We find that a naive presentation of the next-to-leading order power spectra in the original SG shows strange behavior compared with the Newtonian or the relativistic results in the CG, see Figure~\ref{fig:pkSC}.
We then show that such strange behavior is due to the Lagrangian nature of 
the perturbation equations in the SG by introducing the convection terms in both the continuity and the Euler equation. 
With the convection terms, the fluid equations in the SG are 
identical to the Eulerian forms to second order perturbation,
and the equations in the SG coincide with those in the CG to the same order.
The non-linear power spectra in these Eulerian-modified SG equations 
show regularized behaviors, see Figure~\ref{fig:pkSC_LtoE}.
The modification we made in the SG, however, can still be regarded as 
{\it ad hoc}.
In this sense we believe that the SG is not suitable to handle the non-linear power spectra of the density and the velocity perturbations.

Below we begin in Section~\ref{sec:fully-NL} with the basic perturbation equations valid to fully non-linear order in perturbations in both the SG and the CG.
In Section~\ref{sec:third-order} we present hydrodynamic equations valid to third order in both the CG and the SG. We present the relativistic equations using Newtonian hydrodynamic variables, density and velocity perturbations: compare Eqs.\ (\ref{CG-eq1})-(\ref{CG-eq3}) with Eqs.\ (\ref{SG-eq1})-(\ref{SG-eq3}). To linear order the equations in both gauges coincide. To second order, equations in the CG and the SG show relativistic/Newtonian correspondences in the Eulerian form and the Lagrangian form, respectively. The pure Einstein's gravity corrections appearing in the third order in the two gauge conditions are different. In Section~\ref{sec:power-spectra} we present the matter and velocity power spectra obtained by solving the original SG and the Eulerian-modified SG equations. Section~\ref{sec:discussion} is a brief discussion. In the Appendices we present detailed discussion of the gauge issue (Appendix~\ref{sec:GT}), the mode analysis (Appendix~\ref{sec:mode-solutions}), and the Lagrangian modification of the 
equations in the Newtonian theory (Appendix~ \ref{sec:Eulerian-modification}).

\section{Fully non-linear equations}
                                      \label{sec:fully-NL}

We consider the scalar--type perturbations in the flat Friedman background.
Our metric convention follows that of Bardeen in \cite{Bardeen-1988}: 
\begin{equation}
\begin{split}
   \widetilde g_{00} &= - a^2 \left( 1 + 2 \alpha \right) \, ,
   \\
   \widetilde g_{0i} & = - a^2 \beta_{,i} \, ,
   \\
   \widetilde g_{ij} & = a^2 \left[ \left( 1 + 2 \varphi \right) \delta_{ij} + 2 \gamma_{,ij} \right] \, ,
\end{split}
\label{metric-convention}
\end{equation} 
where $x^0 = \eta$ is the conformal time with $c dt \equiv a d \eta$ 
($t$ is the coordinate time). Throughout the paper, we shall use 
$i,j,k,\cdots$ for spatial indices and $a,b,c,\cdots$ for space-time indices.
The inverse metric expanded to third order is presented in Eq.\ (19) of \cite{Hwang-Noh-2007-Third}.
The velocity four-vector is introduced as 
\begin{equation}
   \widetilde u_i \equiv - a v_{,i} \, ,
   \label{v-definition}
\end{equation} 
where we ignored the transverse, vector-type perturbations. 
The other components of the four-vector valid to third order are presented in Eq.\ (22) of \cite{Hwang-Noh-2007-Third}. As will be shown, we have $v = 0$ in both the SG and the CG conditions, thus $\widetilde u_i = 0$ and the four-vector is normal to $\widetilde u_a = \widetilde n_a$. Under this condition ($v \equiv 0$), for a zero-pressure fluid without anisotropic stress, we have 
\begin{equation}
       \widetilde T^0_0 = - \left( \rho + \delta \rho \right) c^2 \, , \quad
       \widetilde T^0_i = 0 \, , \quad
       \widetilde T^i_j = 0 \, .
\end{equation} 
This is true to all perturbation orders. The energy-momentum tensor valid to third order without making any approximation is presented Eq.\ (27) of \cite{Hwang-Noh-2007-Third}. We note that the spatial derivative indices are raised and lowered by $\delta_{ij}$.

The SG takes \cite{Lifshitz-1946,LL} 
\begin{equation}
   \alpha \equiv 0 \quad \text{and} \quad \beta \equiv 0 
\end{equation} 
for the temporal and spatial gauge conditions, respectively. We have $\delta \widetilde g_{00} \equiv 0 \equiv \widetilde g_{0i}$ in the covariant forms, and 
these correspond respectively to $\delta N \equiv 0 \equiv N_\alpha$ in the Arnowitt-Deser-Misner
(ADM) formulation \cite{ADM}; $N (\equiv 1/\sqrt{- \widetilde g^{00}})$ and $N_i( \equiv \widetilde g_{0i})$ are the lapse function and shift vector, respectively. The SG implies the temporal CG condition $\widetilde u_i = 0$ or $\widetilde T^0_i \equiv 0$; see the Appendix B of \cite{Hwang-Noh-2006-SG}. Thus, we have 
\begin{equation}
   v = 0 \, ,
\end{equation} 
and the fluid four-vector becomes normal to $\widetilde u_a = \widetilde n_a$.
To fully non-linear level, we have 
\begin{align}
   \dot \delta & = \left( 1 + \delta \right) \kappa \, ,
   \\
   \dot \kappa + 2 H \kappa
       & = 4 \pi G \rho \delta
       + {1 \over 3} \kappa^2
       + \widetilde \sigma^{ab} \widetilde \sigma_{ab} \, ,
   \\
   \ddot \delta + 2 H \dot \delta - 4 \pi G \rho \delta
      & = 4 \pi G \rho \delta^2
      + {4 \over 3} {\dot \delta^2 \over 1 + \delta}
      + \left( 1 + \delta \right)
      \widetilde \sigma^{ab} \widetilde \sigma_{ab} \, ,
   \label{ddot-delta-eq-SG}
\end{align}
where $\widetilde \sigma_{ab}$ is the shear, 
$\widetilde n^a_{\;\; ;a} \equiv 3H-\kappa$ with $H = \dot{a}/a$.
These are derived in Eqs.\ (18)-(20) in \cite{Hwang-Noh-2006-SG}.
Eq.~(\ref{ddot-delta-eq-SG}) was presented by Kasai in \cite{Kasai-1992} in 
a different form.

The CG takes 
\begin{equation}
   v \equiv 0 \quad \text{and} \quad
       \gamma \equiv 0 
\end{equation} 
as the the temporal and the spatial gauge conditions, respectively. Under these conditions we can show that $\alpha$ vanishes only to linear order.
To fully non-linear level, we have 
precisely the same form of the equations as those in the SG, with the only 
difference being that $\dot\delta$ and $\dot\kappa$ being replaced respectively by
\begin{equation}
   \widehat {\dot \delta} \equiv \dot \delta
       - {1 \over a} \delta_{,i} N^i  \quad \text{and} \quad
       \widehat {\dot \kappa} \equiv \dot \kappa
       - {1 \over a} \kappa_{,i} N^i \, .
\end{equation} 
These are derived in Eqs.\ (14)-(16) in \cite{Hwang-Noh-2006-SG}; $N^i$ expanded to third order is presented in Eq.\ (20) of \cite{Hwang-Noh-2007-Third}.

For $\widetilde u_i \equiv 0$ we have 
\begin{equation}
   \widetilde \sigma^{ab} \widetilde \sigma_{ab}
       = \overline K^i_j \overline K^j_i \, ,
   \label{shear-square}
\end{equation} 
where $\overline K_{ij}$ is the traceless part of extrinsic curvature \cite{Bardeen-1980}; we note that the indices of $N_i$ and $\overline K_{ij}$ are raised and lowered by the ADM three-space metric $h_{ij} \equiv \widetilde g_{ij}$.

We emphasize that the equations and relations up to this point are valid to fully non-linear order. The shear term in Eq.\ (\ref{shear-square}) can be expressed in exact fully non-linear form in the CG \cite{Hwang-Noh-2013}, but such a luxury is not available in the SG. Thus, in the SG we have to expand the term perturbatively.

\section{Third-order perturbations}
                                      \label{sec:third-order}

In this section, we expand the fully non-linear equations in the 
previous section to third order in perturbations.
To have perturbations valid to third order, 
it suffices to expand $\overline K^i_j$ only to second order; this can be found in Eqs.\ (55) and (57) in \cite{Noh-Hwang-2004}.
Considering only the scalar-type perturbations we have 
\begin{align}
   \overline K^i_j \overline K^j_i
       & = {1 \over a^4} \Bigg\{
       \left[ \chi^{,i}{}_{,j} \chi^{,j}{}_{,i}
       - {1 \over 3} \left( \Delta \chi \right)^2 \right]
       \left( 1 - 2 \alpha - 4 \varphi \right)
   \nonumber \\
   & \qquad \quad
       + 2 \chi^{,i}_{\;\;\;j} \left[
       - a \left( \beta_{,i} \varphi^{,j}
       + \beta^{,j} \varphi_{,i}
       + \beta^{,k} \gamma^{,j}{}_{,ik} \right)
       - 2 a^2 \gamma^{,j}{}_{,i} \dot \varphi
       - 2 \gamma^{,jk} \chi_{,ik} \right]
   \nonumber \\
   & \qquad \quad 
       - {2 \over 3} \Delta \chi
       \left[ - a \left( 2 \beta^{,i} \varphi_{,i}
       + \beta^{,i} \Delta \gamma_{,i} \right)
       - 2 a^2 \dot \varphi \Delta \gamma
       - 2 \gamma^{,ij} \chi_{,ij} \right]
       \Bigg\} \, ,
\end{align}
where we set $\chi \equiv a \beta + a^2 \dot \gamma$. In the SG and the CG we set $\beta \equiv 0$ and $\gamma \equiv 0$ (thus $\beta = \chi/a$), respectively. The relation between $\chi$ and $\kappa$ can be derived from the momentum constraint equation. As the $\chi$ terms appear at least in quadratic order we need equation for $\chi$ only to second order. From Eq.\ (69) of \cite{NL-multi-2007} we have 
\begin{align}
   \kappa + {\Delta \over a^2} \chi
       & = {3 \over 2} \Delta^{-1} \nabla^i \Bigg\{
       {1 \over a^2} \left( - \varphi_{,j} \chi^{,j}{}_{,i}
       + {1 \over 3} \varphi_{,i} \Delta \chi
       + {4 \over 3} \varphi \Delta \chi_{,i}
       + \Delta \gamma_{,j} \chi^{,j}{}_{,i}
       + { 1\over 3} \gamma_{,ijk} \chi^{,jk}
       + {4 \over 3} \gamma^{,jk} \chi_{,ijk} \right)
   \nonumber \\
   & \qquad \qquad \quad
       + {1 \over a} \left[ \Delta \beta \varphi_{,i}
       + \beta_{,i} \Delta \varphi
       + {1 \over 3} \left( \beta^{,j} \varphi_{,j} \right)_{,i}
       + \beta^{,jk} \gamma_{,ijk}
       - {1 \over 3} \beta_{,ij} \Delta \gamma^{,j}
       + {2 \over 3} \beta^{,j} \Delta \gamma_{,ij} \right]
   \nonumber \\
   & \qquad \qquad \quad
       + 2 \dot \varphi^{,j} \gamma_{,ij}
       + {4 \over 3} \dot \varphi \Delta \gamma_{,i}
       - {2 \over 3} \dot \varphi_{,i} \Delta \gamma
       \Bigg\}
       \equiv {1 \over a} X \, .
\end{align}

\subsection{Comoving gauge}

By setting $\gamma = 0$ and $\beta = \chi/a$, we have
\begin{align}
   \dot \delta & = \left( 1 + \delta \right) \kappa
       - {1 \over a^2} \delta_{,i} \chi^{,i}
       \left( 1 - 2 \varphi \right) \, ,
   \label{CG-delta-eq} 
   \\
   \dot \kappa + 2 H \kappa
       & = 4 \pi G \mu \delta
       + {1 \over 3} \kappa^2
       - {1 \over a^2} \kappa_{,i} \chi^{,i}
       \left( 1 - 2 \varphi \right)
   \nonumber \\
   & \quad 
       + {1 \over a^4} \left\{
       \left[ \chi^{,ij} \chi_{,ij}
       - {1 \over 3} \left( \Delta \chi \right)^2 \right]
       \left( 1 - 4 \varphi \right)
       - 4 \chi^{,ij} \varphi_{,i} \chi_{,j}
       + {4 \over 3}
       \varphi^{,i} \chi_{,i} \Delta \chi \right\} \, ,
   \label{CG-kappa-eq} 
   \\
   \kappa + {\Delta \over a^2} \chi
       & = 2 \varphi {\Delta \over a^2} \chi
       - {1 \over a^2} \chi^{,i} \varphi_{,i}
       + {3 \over 2 a^2} \Delta^{-1} \nabla^i
       \left( \varphi_{,ij} \chi^{,j}
       + \chi_{,i} \Delta \varphi \right)
       \equiv {1 \over a} X.
\end{align}
Here, we ignore the linear order $\alpha$ term which is already second
order; the momentum conservation equation gives $\alpha = -\chi^{,i} \chi_{,i}/(2 a^2)$ to second order.
We then identify the Newtonian variables as 
\begin{equation}
   \delta \quad \text{and} \quad
       \kappa \equiv - {1 \over a} \nabla \cdot {\bf u}
   \label{kappa-identification}
\end{equation} 
to third order,
and 
\begin{equation}
   \nabla \chi
       = a {\bf u} = a \nabla u
   \label{chi-identification}
\end{equation} 
to linear order. Eqs.~(\ref{CG-delta-eq}) and
(\ref{CG-kappa-eq}) can be arranged as 
\begin{align}
   & \dot \delta
       + {1 \over a} \nabla \cdot {\bf u}
       + {1 \over a} \nabla \cdot \left( \delta {\bf u} \right)
       =
       {1 \over a} \left[ 2 \varphi {\bf u}
       - \nabla \left( \Delta^{-1} X \right) \right]
       \cdot \nabla \delta \, ,
   \label{CG-eq1} 
   \\
   & {1 \over a} \nabla \cdot \left( \dot {\bf u} + H {\bf u} \right)
       + 4 \pi G \mu \delta
       + {1 \over a^2} \nabla \cdot \left( {\bf u} \cdot \nabla {\bf u} \right)
       = - {2 \over 3 a^2} \varphi {\bf u} \cdot \nabla \left( \nabla \cdot {\bf u} \right)
   \nonumber \\
   & \qquad
       + {4 \over a^2} \nabla \cdot \left[
       \varphi \left( {\bf u} \cdot \nabla {\bf u}
       - {1 \over 3} {\bf u} \nabla \cdot {\bf u} \right) \right]
       - {\Delta \over a^2} \left[ {\bf u} \cdot \nabla
       \left( \Delta^{-1} X \right) \right]
       + {1 \over a^2} {\bf u} \cdot \nabla X
       + {2 \over 3 a^2} X \nabla \cdot {\bf u} \, ,
   \label{CG-eq2}
\end{align} 
where 
\begin{equation}
   X \equiv 2 \varphi \nabla \cdot {\bf u}
       - {\bf u} \cdot \nabla \varphi
       + {3 \over 2} \Delta^{-1} \nabla \cdot \left[
       {\bf u} \cdot \nabla \left( \nabla \varphi \right)
       + {\bf u} \Delta \varphi \right] \, .
   \label{CG-eq3}
\end{equation}
Eqs.~(\ref{CG-eq1})-(\ref{CG-eq3}) are
valid to third order in the CG. 
The next-to-leading order matter and velocity power spectra in this 
gauge condition are studied in \cite{JGNH-2011}.

\subsection{Synchronous gauge}

By setting $\alpha = \beta = 0$, we have 
\begin{align}
   \dot \delta & = \left( 1 + \delta \right) \kappa \, ,
   \label{delta-eq-SG} 
   \\
   \dot \kappa + 2 H \kappa
       & = 4 \pi G \mu \delta
       + {1 \over 3} \kappa^2
   \nonumber \\
   & \quad
   \label{kappa-eq-SG}
       + {1 \over a^4} \left[ \chi^{,ij} \chi_{,ij}
       - {1 \over 3} \left( \Delta \chi \right)^2 \right]
       \left( 1 - 4 \varphi \right)
       - {4 \over a^2} \dot \varphi \left[ \chi_{,ij} \gamma^{,ij}
       - {1 \over 3} \left( \Delta \chi \right) \Delta \gamma \right]
       - {4 \over a^4} \chi^{,ij}
       \left[
       \chi_{,jk} \gamma^{,k}{}_{,i}
       - {1 \over 3} \left( \Delta \chi \right)
       \gamma_{,ij} \right] \, ,
   \\
   \kappa + {\Delta \over a^2} \chi
       & = - {3 \over 2 a^2} \varphi^{,i} \chi_{,i}
       + {2 \over a^2} \varphi \Delta \chi
       + {3 \over 2 a^2} \chi^{,i} \Delta \gamma_{,i}
       + {2 \over a^2} \chi_{,ij} \gamma^{,ij}
       + 3 \dot \varphi_{,i} \gamma^{,i}
       + 2 \dot \varphi \Delta \gamma
   \nonumber \\
   & \quad
       - {3 \over 2} \Delta^{-1} \nabla^i \left[
       {1 \over a^2} \varphi_{,i} \Delta \chi
       - {1 \over a^2} \varphi_{,ij} \chi^{,j}
       + \left( {1 \over a^2}
       \chi^{,k} \gamma^{,j}{}_{,ik}
       + 2 \dot \varphi_{,i} \gamma^{,j} \right)_{,j} \right]
       \equiv {1 \over a} X \,.
   \label{chi-SG}
\end{align}

To linear order, considering the SG conditions
$\alpha = \beta = 0$, the basic set of equations is [see Eqs.\ (195)-(201) in \cite{Noh-Hwang-2004}]
\begin{equation}
\begin{split}
   \kappa & = - 3 \dot \varphi - \Delta \dot \gamma \, ,
   \\
   4 \pi G \rho \delta + H \kappa & = - c^2 {\Delta \over a^2} \varphi \, ,
   \\
   \kappa + \Delta \dot \gamma & = 12 \pi G \rho a v \, ,
   \\
   \dot \kappa + 2 H \kappa - 4 \pi G \rho \delta & = 0 \, ,
   \\
   \ddot \gamma + 3 H \dot \gamma - {c^2 \over a^2} \varphi & = 0 \, ,
   \\
   \delta \dot \rho + 3 H \delta \rho & = \rho \kappa \, ,
   \\
   \dot v + H v & = 0 \, .
\end{split}
\end{equation}
We present the gauge transformation properties in the SG in Appendix \ref{sec:GT}.
We can show that $v \propto a^{-1}$ is the gauge mode, and we {\it ignore} it; from Eq.\ (278) in \cite{Noh-Hwang-2004} we have $\hat \alpha = \alpha - (a \xi^0)^\prime/a$ and $\hat v = v - \xi^0$; the SG ($\hat \alpha = 0 = \alpha$) implies $\xi^0 \propto 1/a$, thus the gauge mode behaves as $v_G \propto \xi^0 \propto 1/a$. Thus, we have $\kappa = \dot \delta = - \Delta \dot \gamma$ and $\dot \varphi = 0$, with the solutions 
\begin{equation}
   \delta = c_g t^{2/3} + c_d t^{-1} \, , \quad
       \kappa = \dot \delta = {2 \over 3} c_g t^{-1/3}
       - c_d t^{-2} \, , \quad
       \varphi = - {10 \over 9} \Delta^{-1} c_g \left(a t^{-2/3} \right)^2 \, ,
       \quad
       \gamma = - \Delta^{-1} \delta + \gamma_G \, ,
   \label{SG-sol-linear}
\end{equation} 
where 
$c_g$ and $c_d$ represent the coefficients of the growing and decaying modes respectively, and 
$\gamma_G ({\bf x})$ is a constant gauge mode. In the following we
{\it ignore} this gauge mode.

Using the same identification as in Eqs.\
(\ref{kappa-identification}) and (\ref{chi-identification}), Eqs.\
(\ref{delta-eq-SG})-(\ref{chi-SG}) become 
\begin{align}
   & \dot \delta
       + {1 \over a} \left( 1 + \delta \right) \nabla \cdot {\bf u}
       = 0 \, ,
   \label{SG-eq1} 
   \\
   & {1 \over a} \nabla \cdot \left( \dot {\bf u}
       + H {\bf u} \right) + 4 \pi G \mu \delta
       + {1 \over a^2} u^{,ij} u_{,ij}
       =
       {4 \over a^2} \varphi
       \left[ u^{,ij} u_{,ij}
       - {1 \over 3} \left( \nabla \cdot {\bf u} \right)^2 \right]
   \nonumber \\
   & \qquad
       - {4 \over a^2} u^{,ij} \left[ u_{,jk}
       \left( \Delta^{-1} \delta \right)^{,k}{}_{,i}
       - {1 \over 3} \left( \nabla \cdot {\bf u} \right)
       \left( \Delta^{-1} \delta \right)_{,ij} \right]
       - {2 \over a^2} u^{,ij} \left( \Delta^{-1} X
       \right)_{,ij}
       + {2 \over 3 a^2} \left( \nabla \cdot {\bf u} \right) X \, ,
   \label{SG-eq2}
\end{align} 
where 
\begin{align}
   X &
       = - {3 \over 2} \left( \nabla \varphi \right) \cdot {\bf u}
       + 2 \varphi \nabla \cdot {\bf u}
       - {3 \over 2} {\bf u} \cdot \nabla \delta
       - 2 u^{,ij}
       \Delta^{-1} \delta_{,ij}
   \nonumber \\
   & \quad
       + {3 \over 2} \Delta^{-1} \nabla^i \left\{
       - \left( \nabla_i \varphi \right) \nabla \cdot {\bf u}
       + {\bf u} \cdot \nabla \nabla_i \varphi
       + \nabla^j \left[
       {\bf u} \cdot \nabla \left( \nabla_i \nabla_j
       \Delta^{-1} \delta \right)
       \right] \right\} \, .
   \label{SG-eq3}
\end{align} 
Eqs.~(\ref{SG-eq1})-(\ref{SG-eq3}) are the basic set of
equations to be analyzed to get the next-to-leading order matter and velocity power spectra in the SG.

\section{Power spectra in the synchronous gauge}
                                      \label{sec:power-spectra}

In Fourier space the fluid equations in the SG, 
Eqs.\ (\ref{SG-eq1})-(\ref{SG-eq3}), become
\begin{align}
\label{dot-delta-eq-k}
\dot{\delta}(\bfk,t)
+
\theta(\bfk,t)
= &
-
\int \frac{d^3q_1}{(2\pi)^3}
\int d^3q_2
\delta^{(3)}(\bfk-\bfq_{12})
\delta(\bfq_1,t)
\theta(\bfq_2,t) \, ,
\\
\label{dot-theta-eq-k}
\dot\theta(\bfk,t)
+
2H\theta(\bfk,t)
+
4\pi G\rho \delta(\bfk,t)
= &
-
\int \frac{d^3q_1}{(2\pi)^3}
\int d^3q_2
\delta^{(3)}(\bfk-\bfq_{12})
\theta(\bfq_1,t)
\theta(\bfq_2,t)\frac{(\bfq_1\cdot\bfq_2)^2}{q_1^2q_2^2}
\nonumber
\\
& \hspace{-1.5cm} +4
\int \frac{d^3q_1}{(2\pi)^3}
\int \frac{d^3q_2}{(2\pi)^3}
\int d^3q_3
\delta^{(3)}(\bfk-\bfq_{123})
\left\{
\theta(\bfq_1,t)
\theta(\bfq_2,t)
\varphi(\bfq_3,t)
\left[
\frac{(\bfq_1\cdot\bfq_2)^2}{q_1^2q_2^2}
- {\frac13}
\right]
\right. \nonumber
\\
&\hspace{1.5cm}{-}\left.
\theta(\bfq_1,t)
\theta(\bfq_2,t)
\delta(\bfq_3,t)
\left[
\frac{(\bfq_1\cdot\bfq_2)(\bfq_2\cdot\bfq_3)(\bfq_3\cdot\bfq_1)}{q_1^2q_2^2q_3^2}
-
\frac{1}{3}\frac{(\bfq_1\cdot\bfq_3)^2}{q_1^2q_3^2}
\right]
\right\}
\nonumber
\\
&\hspace{-1.5cm} +
2\int \frac{d^3q_1}{(2\pi)^3}
\int d^3q_2
\delta^{(3)}(\bfk-\bfq_{12})
\theta(\bfq_1,t)
X(\bfq_2,t)
\left[
\frac{1}{3}
-
\frac{(\bfq_1\cdot\bfq_2)^2}{q_1^2q_2^2}
\right] \, ,
\end{align}
where $\theta({\bf x},t) \equiv a^{-1}\nabla\cdot\bfu({\bf x},t)$
is the velocity gradient (or expansion scalar), and 
\begin{align}
\label{X-eq-k}
X(\bfk,t)
& =
\int \frac{d^3q_1}{(2\pi)^3}
\int d^3q_2
\delta^{(3)}(\bfk-\bfq_{12})
\nonumber
\left\{
\theta(\bfq_1,t)\varphi(\bfq_2,t)
\left[
2-\frac{3}{2}\frac{\bfq_1\cdot\bfq_2}{q_1^2}
+\frac{3}{2}\frac{\bfq_{12}\cdot\bfq_2}{q_{12}^2}
\left(
-1+\frac{\bfq_1\cdot\bfq_2}{q_1^2}
\right)
\right]\right.
\\
&\qquad\qquad\qquad\qquad\qquad\qquad\qquad+\left.
\theta(\bfq_1,t)\delta(\bfq_2,t)
\left[
-\frac{3}{2}\frac{\bfq_1\cdot\bfq_2}{q_1^2}
-2\frac{(\bfq_1\cdot\bfq_2)^2}{q_1^2q_2^2}
+
\frac{3}{2}\frac{(\bfq_{12}\cdot\bfq_2)^2(\bfq_1\cdot\bfq_2)}{q_{12}^2q_1^2q_2^2}
\right]
\right\} \, .
\end{align}
Here, we use the shorthand notation that 
$\bfk_{ij\cdots}\equiv\bfk_i+\bfk_j+\cdots$.
In order to calculate the next-to-leading order corrections to the matter
and velocity power spectra in the SG, we solve the equations above 
for $\delta$ and $\theta$ to third order in linear density contrast 
$\delta_1(\bfk)$. 
The details of the calculation are presented in Appendix 
\ref{sec:mode-solutions}.
Although the equations are valid for general cosmology including the cosmological
constant, we find the solutions
for the flat, matter dominated (Einstein-de Sitter, EdS) universe.
In this background, the kernels $F_n$ and $G_n$ take the simplest form 
which respectively relate 
the linear density contrast to 
the $n$-th order non-linear density and velocity fields (see Appendix~\ref{sec:mode-solutions})
Note that as is the case for the CG 
\cite{JGNH-2011}, we find that kernels are quite insensitive to the choice 
of background cosmology.

\begin{figure}[t!]
\centering
\includegraphics*[width=0.49\textwidth]{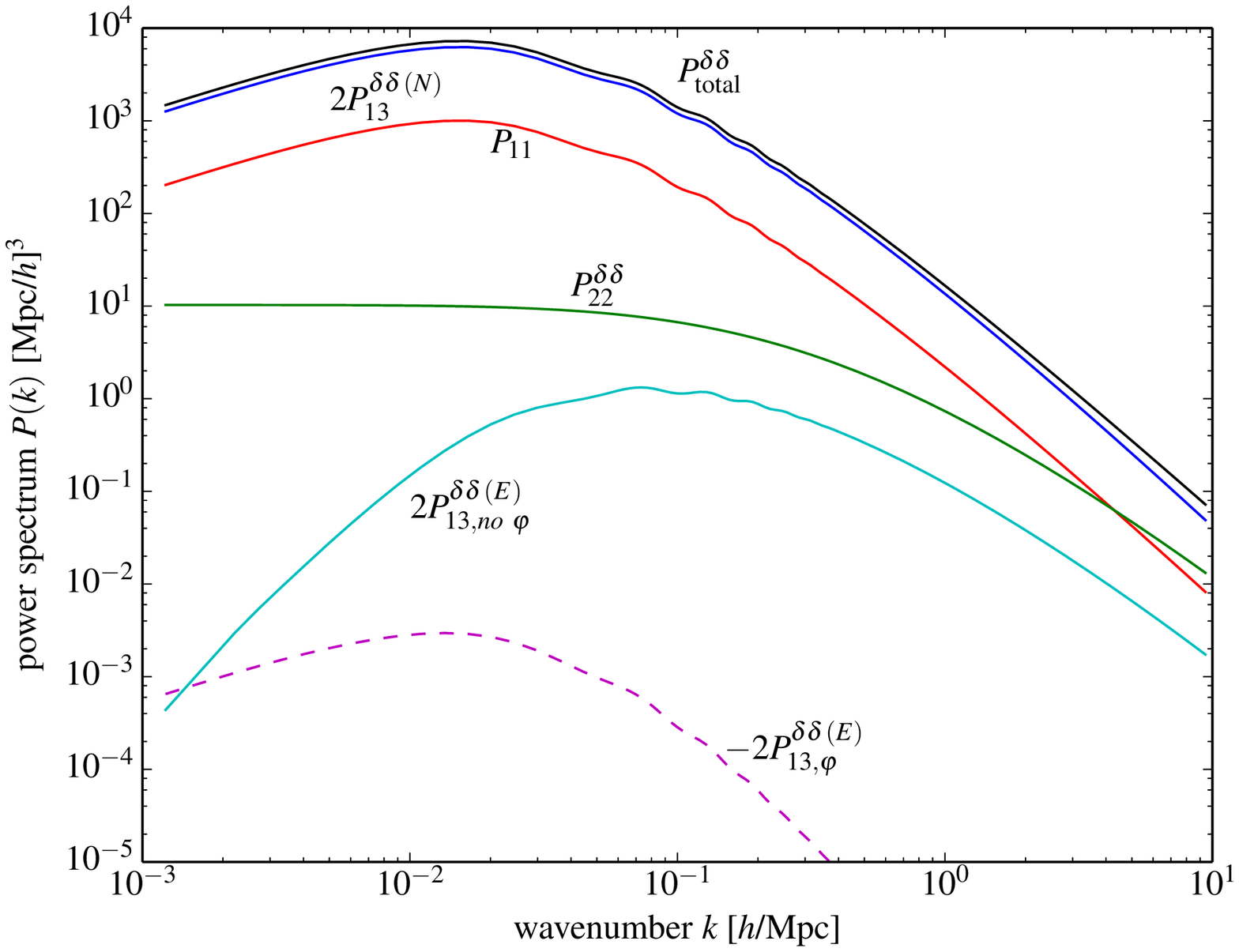}
\includegraphics*[width=0.49\textwidth]{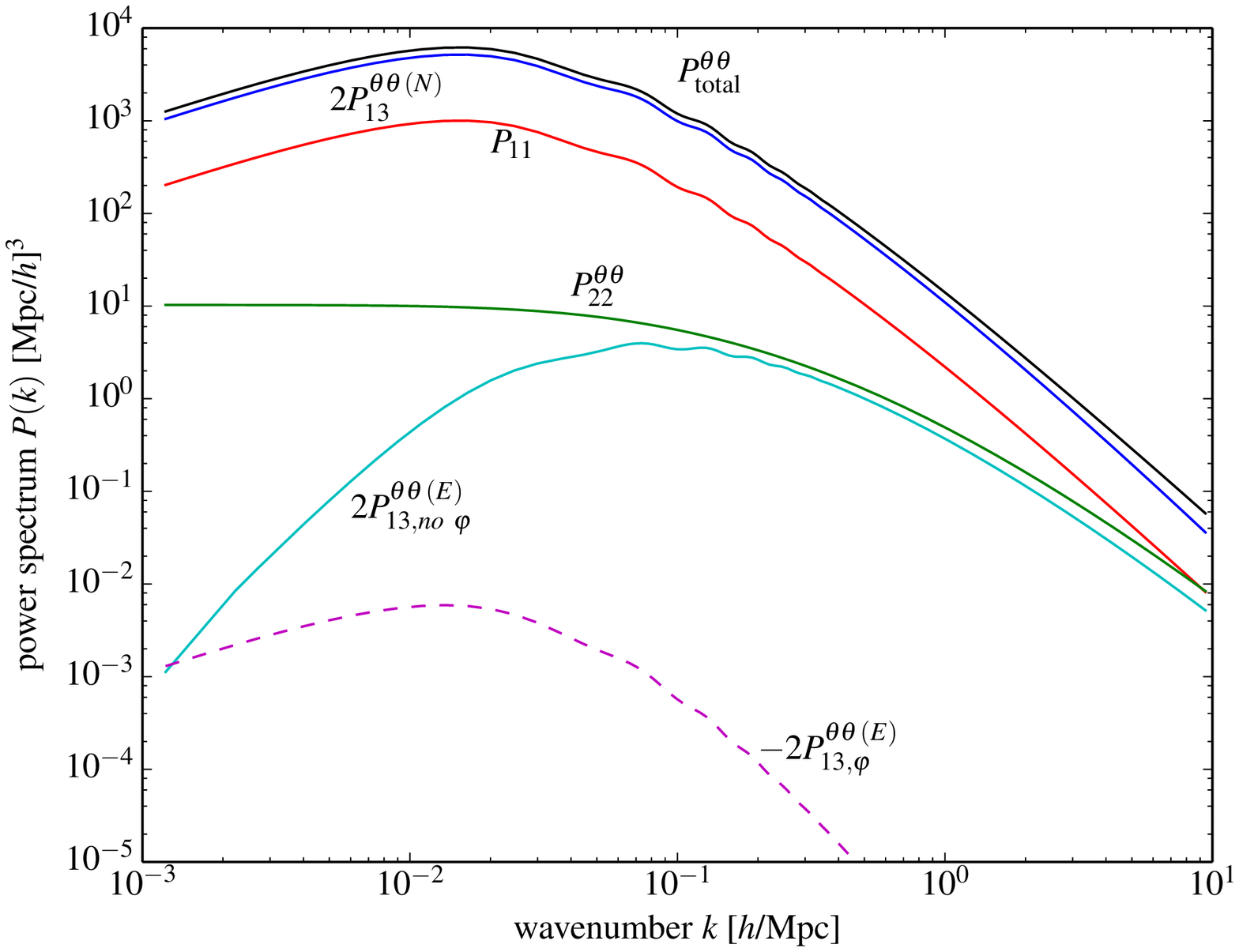}
\caption{
Fully general relativistic total matter (\textit{left panel}) and velocity (\textit{right panel})
power spectra including next-to-leading order corrections
calculated in the SG (black lines)
in EdS universe at redshift $z=6$.
Also shown are the linear matter power spectrum
$P_{11}(k)$ (red lines) and
the next-to leading order corrections to the linear power spectrum:
$P_{22}^{\delta\delta}(k)$ and $P_{22}^{\theta\theta}(k)$ (green lines),
$P_{13}^{\delta\delta\,(N)}(k)$ and $P_{13}^{\theta\theta\,(N)}$ (blue lines),
$P_{13,\mathrm{no}~\varphi}^{\delta\delta\,(E)}(k)$ and 
$P_{13,\mathrm{no}~\varphi}^{\theta\theta\,(E)}(k)$ (cyan lines),
$-P_{13,\varphi}^{\delta\delta\,(E)}(k)$ and 
$-P_{13,\varphi}^{\theta\theta\,(E)}(k)$ (dashed magenta lines).
}
\label{fig:pkSC}
\end{figure}

We find that the second order solutions for density $\delta_2(\bfk)$ 
and velocity gradient $\theta_2(\bfk)$ are
\begin{align}
\label{eq:delta2}
   \delta_2(\bfk,t)
&=
\int \frac{d^3q_1}{(2\pi)^3}
\int d^3q_2
\delta_1(\bfq_1,t)
\delta_1(\bfq_2,t)
\delta^{(3)}(\bfk-\bfq_{12})
F_2(\bfq_1,\bfq_2) \, ,
\\
\label{eq:theta2}
   \theta_2(\bfk,t)
&=
-Hf \int \frac{d^3q_1}{(2\pi)^3}
\int d^3q_2
\delta_1(\bfq_1,t)
\delta_1(\bfq_2,t)
\delta^{(3)}(\bfk-\bfq_{12})
G_2(\bfq_1,\bfq_2) \, ,
\end{align}
with
\begin{align}
F_2(\bfq_1,\bfq_2)
&=
\left[
\frac{5}{7}
+
\frac{2}{7}\frac{(\bfq_1\cdot\bfq_2)^2}{q_1^2q_2^2}
\right] \, ,
\\
G_2(\bfq_1,\bfq_2)
&=
\left[
\frac{3}{7}
+
\frac{4}{7}\frac{(\bfq_1\cdot\bfq_2)^2}{q_1^2q_2^2}
\right] \, ,
\end{align}
being the second order density and velocity kernels in the SG, respectively.
Note that the second order kernels are manifestly symmetric under exchange
of arguments.
The third order solutions are
\begin{align}
\label{eq:delta3}
   \delta_3(\bfk,t)
&=
\int \frac{d^3q_1}{(2\pi)^3}
\int \frac{d^3q_2}{(2\pi)^3}
\int d^3q_3
\delta_1(\bfq_1,t)
\delta_1(\bfq_2,t)
\delta_1(\bfq_3,t)
\delta^{(3)}(\bfk-\bfq_{123})
F_3(\bfq_1,\bfq_2,\bfq_3) \, ,
\\
\label{eq:theta3}
   \theta_3(\bfk,t)
&=
-Hf \int \frac{d^3q_1}{(2\pi)^3}
\int \frac{d^3q_2}{(2\pi)^3}
\int d^3q_3
\delta_1(\bfq_1,t)
\delta_1(\bfq_2,t)
\delta_1(\bfq_3,t)
\delta^{(3)}(\bfk-\bfq_{123})
G_3(\bfq_1,\bfq_2,\bfq_3) \, ,
\end{align}
where
\begin{align}
F_3(\bfq_1,\bfq_2,\bfq_3)
&=
F_3^{(N)}(\bfq_1,\bfq_2,\bfq_3)
+
F_{3,\mathrm{no}~\varphi}^{(E)}(\bfq_1,\bfq_2,\bfq_3)
+
F_{3,\varphi}^{(E)}(\bfq_1,\bfq_2,\bfq_3) \, ,
\label{eq:breakF3}
\\
G_3(\bfq_1,\bfq_2,\bfq_3)
& =
G_3^{(N)}(\bfq_1,\bfq_2,\bfq_3)
+
G_{3,\mathrm{no}~\varphi}^{(E)}(\bfq_1,\bfq_2,\bfq_3)
+
G_{3,\varphi}^{(E)}(\bfq_1,\bfq_2,\bfq_3) \, ,
\label{eq:breakG3}
\end{align}
are, respectively, the third order SG kernels for density field and 
velocity gradient field. All kernels are symmetrized under exchange of 
arguments.
For later convenience, 
in Eqs.~\eqref{eq:breakF3} and \eqref{eq:breakG3} 
we separate the kernels as 
Newtonian and purely relativistic parts, 
denoted by the superscripts $(N)$ and $(E)$ respectively.
We identify the Newtonian kernels as the solution for the equations 
truncated at second order
[the first lines of Eqs.~(\ref{dot-delta-eq-k}) and 
(\ref{dot-theta-eq-k})], which coincide with the usual fluid equations in
the Lagrangian coordinate as shown in Appendix \ref{sec:Eulerian-modification}.
Finally, unlike the density and velocity kernels in CG
\cite{JGNH-2011}, there are terms in the relativistic kernels which are not 
proportional to the gravitational potential $\varphi$: 
we further separate the relativistic kernels as 
$F^{(E)}_{3, \mathrm{no}~\varphi}$, 
$F^{(E)}_{3, \varphi}$
and 
$G^{(E)}_{3, \mathrm{no}~\varphi}$, 
$G^{(E)}_{3, \varphi}$.
We present the explicit expressions for all of the third order kernels in 
Appendix \ref{sec:mode-solutions}.

From the non-linear solutions up to third order, we calculate the non-linear 
power spectra of density and velocity as 
\begin{equation}
P(k,z)
=
D^2(z)P_{11}(k)
+
D^4(z)
\left[
P_{22}(k) + 2 P_{13}(k)
\right] \, ,
\label{eq:pkSG_division}
\end{equation}
where $D(z)$ is the linear growth function.
We define $P_{ab}^{XY}(k)$ is the non-linear correction to the
matter/velocity power spectrum. For the matter, it is defined by
\begin{equation}
\left<
\delta_a(\bfk)\delta_b(\bfk')
\right>
\equiv
(2\pi)^3 \delta^{(3)}(\bfk+\bfk') P_{ab}^{\delta\delta}(k).
\end{equation}
In order to match the amplitude of the velocity-gredient power sepctrum 
on large scales to the linear matter power spectrum $P_{11}(k)$, 
we define $P_{ab}^{\theta\theta}$ as
\be
\left<\theta_a(\bfk)\theta_b(\bfk')\right>
=
(2\pi)^3 \delta^{(3)}(\bfk+\bfk')
(Hf)^2 P_{ab}^{\theta\theta}(k)
\ee
Note that we assume the statistical isotropy of the primordial universe 
which is retained in non-linear order and removes the angular dependence 
of the power spectrum $P_{ab}^{XY}$.
By using the non-linear solutions in Eqs.~(\ref{eq:delta2}) and (\ref{eq:theta2}), we find
$P_{22}^{\delta\delta}(k)$ and
$P_{22}^{\theta\theta}(k)$ as,
with $x \equiv \cos\theta$,
\begin{align}
P_{22}^{\delta\delta}(k)
& =
\frac{k^3}{2\pi^2}
\int
dr
\int_{-1}^1 dx
P_{11}(kr)
P_{11}\left( k\sqrt{1+r^2-2rx} \right)
\left[
\frac{r(5+7r^2-14rx+2x^2)}{7(1+r^2-2rx)}
\right]^2 \, ,
\\
P_{22}^{\theta\theta}(k)
& =
\frac{k^3}{2\pi^2}
\int
dr
\int_{-1}^1 dx
P_{11}(kr)
P_{11}\left( k\sqrt{1+r^2-2rx} \right)
\left[
\frac{r(3+7r^2-14rx+4x^2)}{7(1+r^2-2rx)}
\right]^2 \, ,
\end{align}
where $r$ and $x$ are the magnitude of dummy integration momentum $\bfq$ and the cosine between $\bfq$ and $\bfk$ respectively, i.e. $q \equiv kr$ and $\bfk\cdot\bfq \equiv k^2rx$.
We divide $P_{13}^{XY}(k)$ by three parts according the three pieces of the 
third order kernels in Eqs. (\ref{eq:breakF3}) and (\ref{eq:breakG3}):
\begin{align}
P_{13}^{XY}(k)
=
P_{13}^{XY\,(N)}(k)
+
P_{13,\mathrm{no}~\varphi}^{XY\,(E)}(k)
+
P_{13,\varphi}^{XY\,(E)}(k) \, ,
\end{align}
where for density power spectrum,
\begin{equation}
\begin{split}
P_{13}^{\delta\delta\,(N)}(k)
& =
\frac{k^3P_{11}(k)}{(2\pi)^2}
\int dr P_{11}(kr)
\frac{1}{378r^3}
\left[
-4r\left(3 + 10r^2 - 413r^4 + 3r^6\right)
-6\left(r^2-1\right)^2
\left(1 + 5r^2 + r^4\right)
\log\left|\frac{1-r}{1+r}\right|
\right] \, ,
\\
P_{13,\mathrm{no}~\varphi}^{\delta\delta\,(E)}(k)
& =
\frac{k^3P_{11}(k)}{(2\pi)^2}
\int dr P_{11}(kr)
\frac{1}{864r^3}
\left[
4r\left(-21 + 53r^2 - 13r^4 + 3r^6\right)
+6\left(r^2-1\right)^3
\left(7+r^2\right)
\log\left|\frac{1-r}{1+r}\right|
\right] \, ,
\\
P_{13,\varphi}^{\delta\delta\,(E)}(k)
& = \frac{k^3P_{11}(k)}{(2\pi)^2}
\left(\frac{k_H}{k}\right)^2
\int dr P_{11}(kr)\frac{1}{168}
\left[
\frac{20}{r^2}\left(17r^4+54r^2-3\right)
-
\frac{30}{r^3}\left(r^2-1\right)^2\left(1+5r^2\right)
\log\left|\frac{1-r}{1+r}\right|
\right] \, ,
\end{split}
\end{equation}
and for velocity power spectrum,
\begin{equation}
\begin{split}
P_{13}^{\theta\theta\,(N)}(k)
& =
\frac{k^3P_{11}(k)}{(2\pi)^2}
\int dr P_{11}(kr)
\frac{1}{126r^3}
\left[
-4r\left(3 + 10r^2 - 107r^4 + 3r^6\right)
-6\left(r^2-1\right)^2
\left(1 + 5r^2 + r^4\right)
\log\left|\frac{1-r}{1+r}\right|
\right] \, ,
\\
P_{13,\mathrm{no}~\varphi}^{\theta\theta\,(E)}(k)
& =
3 P_{13,\mathrm{no}~\varphi}^{\delta\delta\,(E)}(k) \, ,
\\
P_{13,\varphi}^{\theta\theta\,(E)}(k)
& =
2 P_{13,\varphi}^{\delta\delta\,(E)}(k) \, .
\end{split}
\end{equation}
As expected, only 
$P_{13,\varphi}^{\delta\delta(E)}$ and
$P_{13,\varphi}^{\theta\theta(E)}$ depend 
on $k_H=aH/c$, the wavenumber corresponding to the comoving horizon.

We show the results of the numerical integrations in Figure~\ref{fig:pkSC}.
We evaluate the non-linear power spectra by using the linear matter 
power spectrum at $z=6$.
Left and right panels in Figure~\ref{fig:pkSC} show, respectively, 
the matter power spectrum and the velocity power spectrum in the SG.
For each panels, top black curve shows the full calculation of the 
next-to-leading order power spectrum, and the red curve shows the linear
power spectrum. As clearly shown, the next-to-leading order
power spectra of matter and velocity overwhelm the linear ones
in the SG, which indicates the breakdown of the perturbation theory scheme 
itself in the SG, or inappropriate nature of the constant-time hypersurface 
(in particular the spatial coordinate and the Fourier wavenumber)
in the SG. For the further discussion, see the next section and 
Appendix \ref{sec:Eulerian-modification}.

In order to further investigate the problem, we show each contribution in 
Eq.\ \eqref{eq:pkSG_division}
separately in the same figure:
$P_{22}^{\delta\delta}(k)$ and $P_{22}^{\theta\theta}(k)$ (green line),
$2P_{13}^{\delta\delta\,(N)}(k)$ and $2P_{13}^{\theta\theta\,(N)}$
(blue line),
$2P_{13,\mathrm{no}~\varphi}^{\delta\delta\,(E)}(k)$ and 
$2P_{13,\mathrm{no}~\varphi}^{\theta\theta\,(E)}(k)$ (cyan line) and
$-2P_{13,\varphi}^{\delta\delta\,(E)}(k)$ and 
$-2P_{13,\varphi}^{\theta\theta\,(E)}(k)$ (dashed magenta line).
Note that the $P_{13,\varphi}^{XX\,(E)}(k)$ terms are negative for both cases,
and we show the absolute values as dashed lines.
Except for these small contributions from the pure Einstein's gravity term
with $\varphi$, which is highly suppressed in all 
scales due to the smallness of $\varphi$, all the other terms 
contribute significantly compared to the linear power spectrum. 
First of all, the $P_{13}^{XX\,(N)}$ terms 
dominate over the linear power spectrum on all scales. 
The $P_{22}^{XX\,(N)}$ terms exceed the linear power spectrum on very-large and
small scales, and the $P_{13,\text{no }\varphi}^{XX\,(E)}$ terms exceed the linear 
power spectrum on small scales.

The two terms $P_{22}^{XX\,(N)}$ and $P_{13}^{XX\,(N)}$, which exceed 
the linear power spectrum on large scales, are both \textit{Newtonian} in a 
sense that they are the non-linear solutions of the Newtonian fluid equations 
in the Lagrangian form.
On large scale $k\to 0$ limit, the $P_{22}^{XX}(k)$ approaches constant:
\begin{equation}
P_{22}^{XX}(k\to0)
= 2\int \frac{d^3q}{(2\pi)^3}
P_{11}^2(q) \, ,
\end{equation}
as $F_2(\bfq,-\bfq)=G_2(\bfq,-\bfq)= 1$. Therefore,
on sufficiently large scales, $P_{22}^{XX}(k)$ must exceed $P_{11}(k)$, 
which monotonically decreases toward large scales ($k\to 0$).
In the same limit, the third order kernels asymptote to 
\begin{align}
F_3^{(N)}(\bfq,-\bfq,\bfk\to 0) & = \frac{13}{21} + \frac{8}{21}\mu^2 \, ,
\\
G_3^{(N)}(\bfq,-\bfq,\bfk\to 0) & = \frac{3}{7} + \frac{4}{7}\mu^2 \, ,
\end{align}
with $\mu \equiv \bfk\cdot\bfq/(kq)$,
then we find
\begin{align}
P_{13}^{\delta\delta\,(N)}(k\to 0) & = 
\frac{47}{21}P_{11}(k)\int \frac{d^3q}{(2\pi)^3}P_{11}(q)
\equiv 
\frac{47}{21}\sigma^2 P_{11}(k),
\\
P_{13}^{\theta\theta\,(N)}(k\to 0) & =
\frac{13}{7}\sigma^2 P_{11}(k) \, .
\end{align}
Here, we define the root mean square of the linear matter fluctuation $\sigma$,
which diverges for the linear matter power spectrum in the standard $\Lambda$CDM.
Therefore, formally the $P_{13}^{XX\,(N)}$ terms must diverge as well.
In other words, the result shown here depends on the small-scale cutoff of the
integration. For the presentation purpose, we explicitly set the smallest scale
wavenumber (upper bound of integration) to be $k_\text{max} = 5~h/\mathrm{Mpc}$ in 
Figure~\ref{fig:pkSC}.

Finally, it is worth noting that, in contrast to the CG case \cite{JGNH-2011}
where relativistic corrections are suppressed on all scales,
$P_{13,\mathrm{no}\,\varphi}^{XX}(k)$ 
contributes to the small scale non-linear power spectrum. 
It is because this term is defined as the part of the third order solution 
which does not explicitly contain $\varphi$. The existence of such terms 
should not be surprising as the non-linear gauge transformation 
(see Appendix \ref{sec:GT}) do contain terms without $\varphi$. 
Thus, even though the third order solution of pure Einstein's gravity 
in the CG consists only of terms including $\varphi$, solutions
in other general gauges must contain terms which do not explicitly contain 
$\varphi$.

\section{Convective derivative interpretation of the synchronous gauge time derivative}
\begin{figure}[t!]
\centering
\includegraphics*[width=0.49\textwidth]{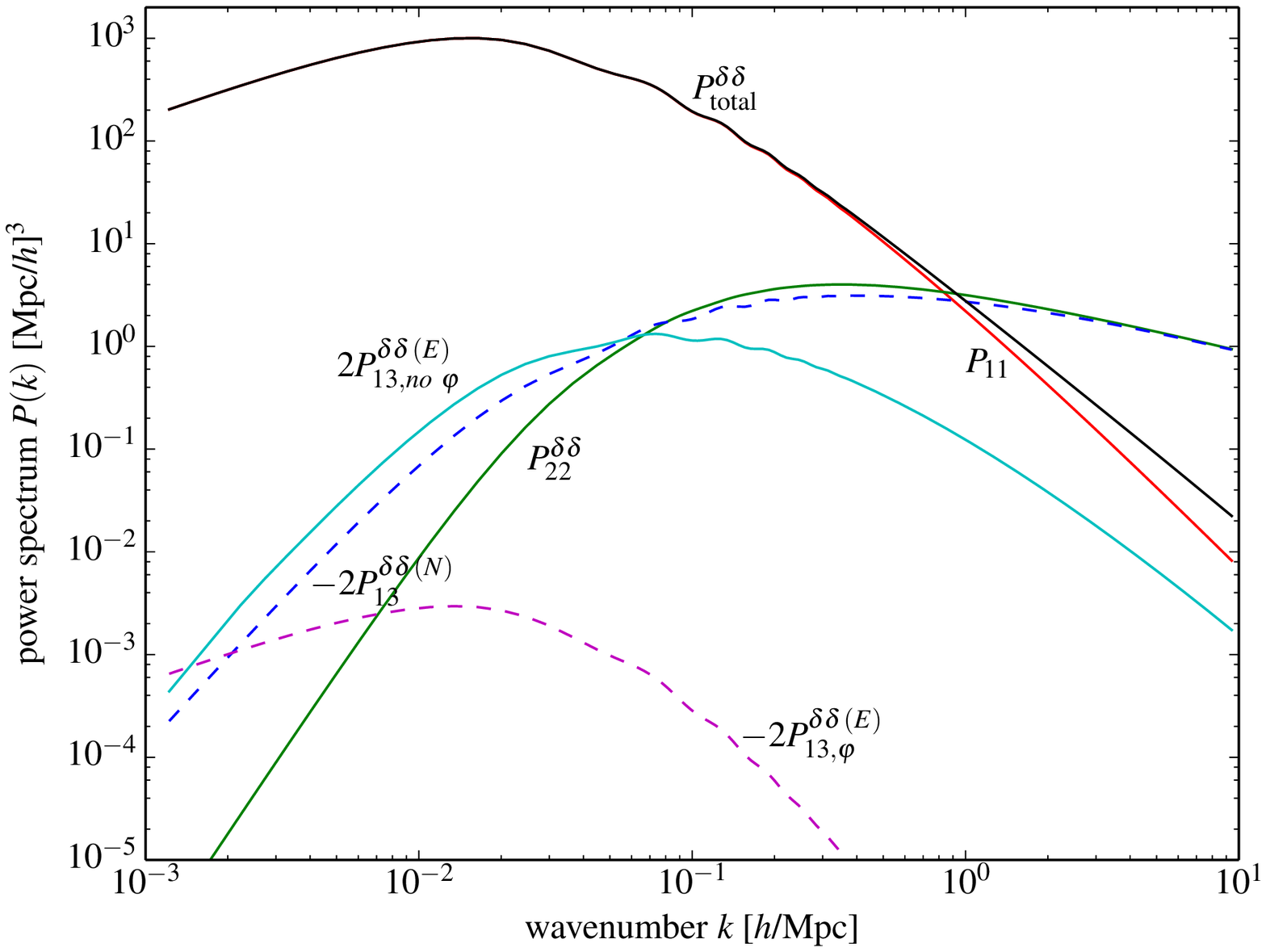}
\includegraphics*[width=0.49\textwidth]{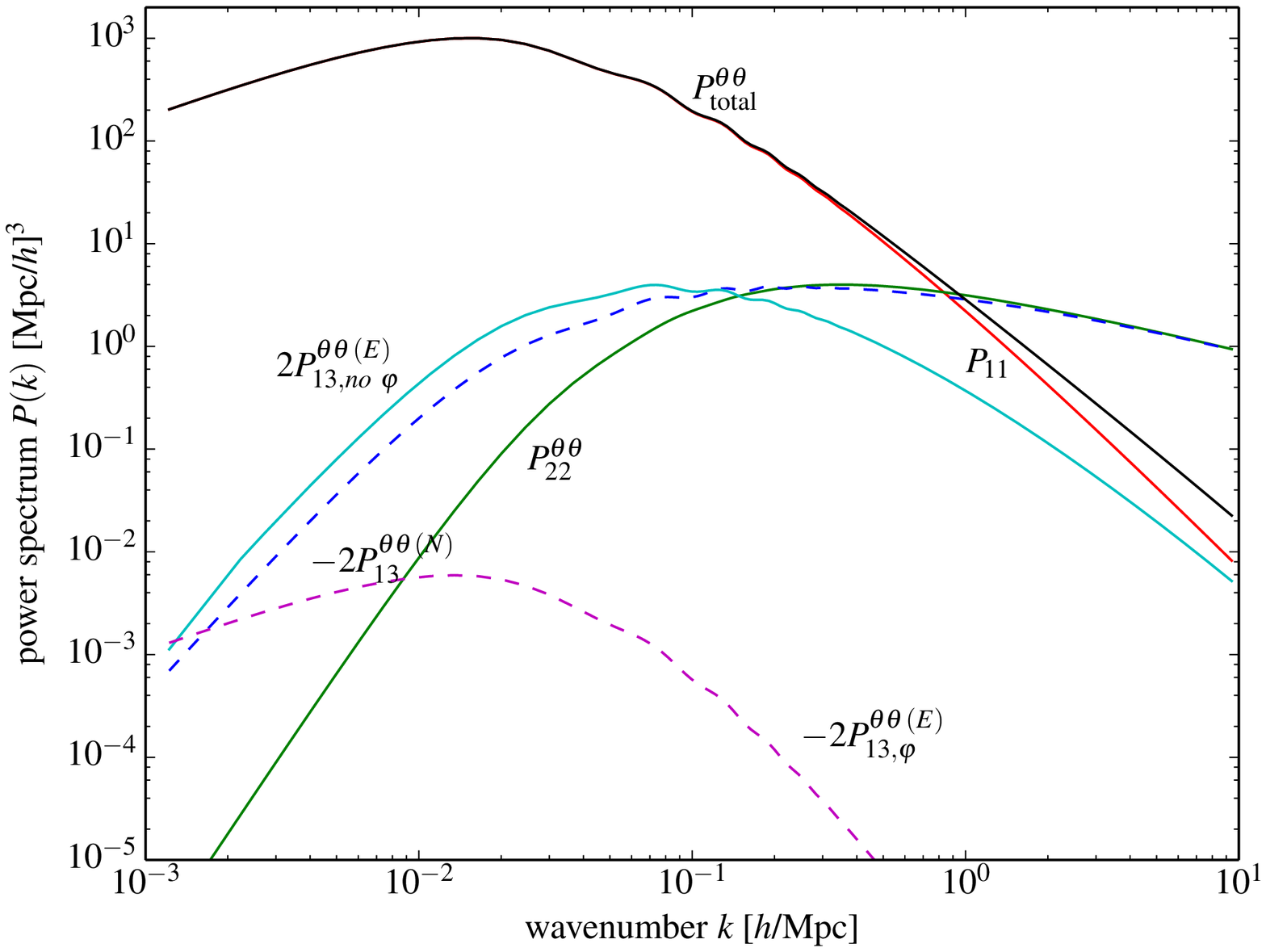}
\caption{
Same as Figure~\ref{fig:pkSC}, but after the \textit{ad-hoc}
Lagrangian to Eulerian transformation by interpreting the time derivative
in the SG as a convective derivative.
Line symbols and colors are the same as Figure~\ref{fig:pkSC}.
Note that after the transformation, 
$P_{13}^{\delta\delta(N)}(k)$ and $P_{13}^{\theta\theta(N)}(k)$ 
(blue, dashed line) become negative.
}
\label{fig:pkSC_LtoE}
\end{figure}

In the previous section, we show that the non-linear corrections to the 
matter and the velocity power spectra in the SG have
serious problems. In particular, the $P_{13}^{(N)}$ terms, which formally diverge,
are larger than $P_{11}$ on all scales even with the moderate
cutoff at $k_{\rm max} = 5~h/\mathrm{Mpc}$. 
This is a clear indication of the breakdown of the perturbation theory in the SG.
That is, the SG is not appropriate for the perturbative analysis of the 
non-linear matter/velocity evolution. 
Such a bad behavior of the matter/velocity power spectrum in the SG is 
expected because the time coordinate in the SG is defined along the
trajectory of each particle. In other words, when interpreting with the 
Newtonian language, the SG corresponds
to the ``Lagrangian'' view in the sense that is usually used in fluid dynamics.

To make this point more clear, we shall solve the fluid equations in 
the SG once more, but with different interpretation of the ``time'' coordinate. 
That is, we interpret the time derivative in the fluid equations in the SG  
as a convective time derivative:
\begin{equation}
\frac{d}{dt} 
\to
\frac{d}{dt} + \frac1a\bf{u}\cdot\nabla.
\end{equation}
Under this transformation, the fluid equations become
\begin{align}
&\dot{\delta} + \frac{1}{a}(1+\delta)\nabla\cdot{\bf u}=0
\to\;
\dot{\delta}+\frac1a\nabla\cdot\left[(1+\delta){\bf u}\right] 
=0 \, ,
\\
&\frac{1}{a}
\nabla\cdot\left(\dot{\bf u}+H{\bf u}\right)
+4\pi G \rho\delta
+\frac{1}{a^2}u^{,ij}u_{,ij}
= [\mathrm{RHS~of~Eq.~} (\ref{SG-eq2})]
\vs
\to\;&
\dot\theta
+
2H \theta
+ 4\pi G \rho\delta
+
\frac{1}{a^2}
\nabla\cdot\left[\left({\bf u}\cdot\nabla\right){\bf u}\right]
= [\mathrm{RHS~of~Eq.~} (\ref{SG-eq2})] \, .
\end{align}
Note that the left hand sides of above equations coincide with Newtonian fluid
equations or fluid equations in the CG.
Then, the Fourier space continuity and Euler equations 
in this case are
\begin{align}
\dot\delta(\bfk,t)
+
\theta(\bfk,t)
& =
-\int \frac{d^3q_1}{(2\pi)^3}
\int d^3q_2
\delta^{(3)}(\bfk-\bfq_{12})
\frac{\bfk\cdot\bfq_2}{q_2^2}
\delta(\bfq_1,t)\theta(\bfq_2,t) \, ,
\\
\dot\theta(\bfk,t)
+
2H\theta(\bfk,t)
+
\frac{3}{2} H^2\Omega_m\delta(\bfk,t)
=&
-
\int \frac{d^3q_1}{(2\pi)^3}
\int d^3q_2
\delta^{(3)}(\bfk-\bfq_{12})
\frac{k^2(\bfq_1\cdot\bfq_2)}{2q_1^2q_2^2}
\theta(\bfq_1,t)\theta(\bfq_2,t)
\vs
&\hspace{-1.5cm}+4
\int \frac{d^3q_1}{(2\pi)^3}
\int \frac{d^3q_2}{(2\pi)^3}
\int d^3q_3
\delta^{(3)}(\bfk-\bfq_{123})
\left\{
\theta(\bfq_1,t)
\theta(\bfq_2,t)
\varphi(\bfq_3,t)
\left[
\frac{(\bfq_1\cdot\bfq_2)^2}{q_1^2q_2^2}
-\frac13
\right]
\right. \nonumber
\\
&\hspace{1.5cm}{-}\left.
\theta(\bfq_1,t)
\theta(\bfq_2,t)
\delta(\bfq_3,t)
\left[
\frac{(\bfq_1\cdot\bfq_2)(\bfq_2\cdot\bfq_3)(\bfq_3\cdot\bfq_1)}{q_1^2q_2^2q_3^2}
-
\frac{1}{3}\frac{(\bfq_1\cdot\bfq_3)^2}{q_1^2q_3^2}
\right]
\right\}
\nonumber
\\
&\hspace{-1.5cm}+2
\int \frac{d^3q_1}{(2\pi)^3}
\int d^3q_2
\delta^{(3)}(\bfk-\bfq_{12})
\theta(\bfq_1,t)
X(\bfq_2,t)
\left[
\frac{1}{3}
-
\frac{(\bfq_1\cdot\bfq_2)^2}{q_1^2q_2^2}
\right].
\end{align}
To second order in perturbation, the equations above exactly coincide with
those in the CG. That is, we, again, recover the 
Newtonian-relativistic correspondence. 
The equations for the third order kernels are 
\begin{align}
& \frac{1}{H}\frac{dF_3}{dt}
+ 3 F_3 - G_3
=
F_2(\bfq_1,\bfq_3)\frac{\bfk\cdot\bfq_2}{q_2^2}
+
G_2(\bfq_2,\bfq_3)\frac{\bfk\cdot\bfq_{23}}{q_{23}^2} \, ,
\\
\nonumber
&\frac{1}{H}\frac{dG_3}{dt} + 2 G_3 - \frac{3}{2}(F_3-G_3)
\\
\nonumber
=&
\Biggl\{
\frac{k^2(\bfq_1\cdot\bfq_{23})}{q_1^2q_{23}^2}G_2(\bfq_2,\bfq_3)
{+}
4
\left[
\frac{(\bfq_1\cdot\bfq_2)(\bfq_2\cdot\bfq_3)(\bfq_3\cdot\bfq_1)}{q_1^2q_2^2q_3^2}
-
\frac{1}{3}\frac{(\bfq_1\cdot\bfq_3)^2}{q_1^2q_3^2}
\right]
\vs
&+
\left[
3\frac{\bfq_2\cdot\bfq_3}{q_2^2}
+
4\frac{(\bfq_2\cdot\bfq_3)^2}{q_2^2q_3^2}
-
3\frac{(\bfq_{23}\cdot\bfq_3)^2(\bfq_2\cdot\bfq_3)}{q_{23}^2q_2^2q_3^2}
\right]
\left[
\frac{1}{3}
-
\frac{(\bfq_1\cdot\bfq_{23})^2}{q_1^2q_{23}^2}
\right]
\Biggl\}
\vs
&-
\left\{
5\frac{k_H^2}{q_3^2}
\left[
2-\frac{3}{2}\frac{\bfq_2\cdot\bfq_3}{q_2^2}
+\frac{3}{2}\frac{\bfq_{23}\cdot\bfq_3}{q_{23}^2}
\left(
-1+\frac{\bfq_2\cdot\bfq_3}{q_2^2}
\right)
\right]
\left[
\frac{1}{3}
-
\frac{(\bfq_1\cdot\bfq_{23})^2}{q_1^2q_{23}^2}
\right]
+
10\frac{k_H^2}{q_3^2}
\left[
\frac{(\bfq_1\cdot\bfq_2)^2}{q_1^2q_2^2}
-{\frac13}
\right]
\right\} \, .
\end{align}
Solving these equations yields the same relativistic kernels 
as what we found in Section~\ref{sec:power-spectra}, and 
the Newtonian kernels are identical to those in the {\em Eulerian} frame (see Appendix~\ref{sec:Eulerian-modification}).
Thus, under the time derivative transformation, 
the $P_{22}^{XX}$ and 
the $P_{13}^{XX\,(N)}$ terms coincide exactly with their Newtonian
counterparts.

We show the results of the numerical integrations in Figure~\ref{fig:pkSC_LtoE}.
Here, again, we evaluate the non-linear power spectrum at $z=6$
and show the matter power spectrum (left) and the velocity power spectrum 
(right).
For each panel, top black curve shows the full calculation of the 
next-to-leading order power spectrum, and the red curve shows the linear
power spectrum. After replacing the problematic $P_{22}^{XX}$ and 
$P_{13}^{XX\,(N)}$ terms to the corresponding functions in the Newtonian 
non-linear perturbation theory, the non-linear power spectra show 
a regularized behavior for both matter and velocity.

Based on this analyses we conclude that the strange behaviors of the 
non-linear power spectra in the SG in Figure \ref{fig:pkSC} are not due to some 
pathological breakdown of the perturbation theory in that gauge, but
because of the Lagrangian nature of the time coordinate in the SG.
The wavenumbers used in the SG in Section~\ref{sec:power-spectra} are not the 
one based on the Eulerian coordinate, thus are inappropriate.
This point was suggested by Professor Masumi Kasai in a private discussion.
We further clarify this by showing the non-linear power spectra based on the
Lagrangian coordinate in the Newtonian context: see Appendix 
\ref{sec:Eulerian-modification} and Figure 
\ref{fig:pk_Newtonian}.

\section{Discussion}
                                      \label{sec:discussion}
In this paper, we show that the synchronous gauge (SG) is an inappropriate 
gauge to describe the non-linear evolution of the density and velocity fields.
It is because the time coordinate of the SG follows the trajectories of each
particle; thus it corresponds to the Lagrangian picture of the Newtonian fluid
dynamics. This should be contrasted with the comoving gauge (CG) whose 
Newtonian correspondence is the Eulerian picture of the fluid dynamics.

As a result, the next-to-leading order correction terms formally diverge if 
linear matter power spectrum is extended to the infinitely small scale and
depend on the cutoff imposed there. 
By closely examining each term in the solution, we find that 
it is the \textit{Newtonian} terms that cause the problematic behavior.
Namely, the \textit{Newtonian part} of the continuity and the Euler 
equations in the SG correspond to the Newtonian fluid dynamics equations in 
the Lagrangian coordinate. We interpret this correspondence as the result of 
the time coordinate of the SG coinciding with the worldline of a comoving 
observer; such a time coordinate is closely related to the \textit{convective 
time} coordinate in the Lagrangian picture of fluid dynamics. 
We further justify the interpretation by trading all the time dependence 
in the resulting fluid equations to the convective time, and show that the 
next-to-leading order power spectra are well regulated with the convective time
derivative. After the time coordinate transformation, like the case for the 
CG, the relativistic correction terms are suppressed on 
large scales, but we also observe an interesting rise of the no-$\varphi$ contribution
on scales smaller than a few $\times 0.1~h/\mathrm{Mpc}$.

Recent studies~\cite{Baldauf:2011bh,Jeong:2011as} have shown that the linear galaxy bias, 
which is determined by the halo/galaxy formation physics on scales 
smaller than the Hubble horizon ($r<c/H$), can be extended to the near horizon 
scales only in the SG. It is because, in the SG, the constant-coordinate-time 
hypersurface coincides with the constant-proper-time hypersurface and all of 
the local cosmic events are synchronized by the coordinate 
time~\cite{Jeong-2013}. At linear order, density contrast in the SG coincides with that
in the CG and its power spectrum is the same as the usual linear matter 
power spectrum. Our calculation, however, implies that we cannot simply extend 
the statement to higher order, because of the inappropriate nature of the 
Lagrangian coordinate embedded in the SG analysis.
As a result, the power spectra of the non-linear density and 
velocity fields in the SG are not well regulated, and lead to the infinite 
corrections to the linear power spectra.
One possible remedy of the situation is to consider the ``observed'' galaxy
power spectrum which must be calculated from the observed angular position 
and redshift of galaxies. Because of the light deflection effect and time 
delay, the observed position and the coordinate position of galaxies are 
systematically different.
\cite{Jeong:2011as,obsPg}
studied these effects to linear order in perturbation, and 
\cite{Bartolo:2010ec} partially included the second order effects.
Based on the success of those calculations, we surmise that the same may be
true for second order so that the second order projection effects 
cancel out the non-linear divergence that we find in the paper. 
We leave the detailed calculations as a future work.

%
%
\subsection*{Acknowledgments}

We wish to thank Professors Masumi Kasai and Misao Sasaki for useful 
discussion. We thank the Topical Research Program ``Theories and practices 
in large scale structure formation'', supported by the Asia Pacific Center 
for Theoretical Physics, while this work was under progress.
J.H.\ was supported by Basic Science Research Program through the National Research Foundation (NRF) of Korea funded by the Ministry of Science, ICT and future Planning (No.\ 2013R1A2A2A01068519). H.N.\ was supported by NRF of Korea funded by the Korean Government (No.\ 2012R1A1A2038497).
J.G. acknowledges the Max-Planck-Gesellschaft, the Korea Ministry of Education, Science and Technology, Gyeongsangbuk-Do and Pohang City for the support of the Independent Junior Research Group at the Asia Pacific Center for Theoretical Physics. JG is also supported by a Starting Grant through the Basic Science Research Program of the National Research Foundation of Korea (2013R1A1A1006701).

%
%
\appendix

\section{Gauge transformation}
                                      \label{sec:GT}

We consider the gauge transformation $\hat x^a = x^a + \widetilde \xi^a (x^e)$.
We can show that the gauge transformation of $\widetilde T^0_i$ involves $\widetilde \xi^0_{\;\;,i}$
in all terms to third order. As we have $v = 0$ in the two gauge conditions we are considering, we have $\widetilde \xi^0 = 0$ for both and even under gauge transformation between them. By setting $\widetilde \xi^0 = 0$, the remaining gauge transformation involves only the spatial one $\widetilde \xi^i \equiv \xi^i$ whose index is raised and lowered by $\delta_{ij}$, and for the scalar-type perturbations we may set $\xi_i \equiv \xi_{,i}$. We have 
\begin{align}
   \hat \alpha
       &= \alpha
       - \alpha_{,i} \xi^{,i}
       - a \beta_{,i} \dot \xi^{,i}
       - {1 \over 2} a^2 \dot \xi^{,i} \dot \xi_{,i}
       + \alpha_{,i} \xi^{,i}{}_{,j} \xi^{,j}
       + {1 \over 2} \alpha_{,ij} \xi^{,i} \xi^{,j}
       + a \beta_{,i} \xi^{,i}{}_{,j} \dot \xi^{,j}
       + a \left( \beta_{,i} \dot \xi^{,i} \right)_{,j} \xi^{,j}
   \nonumber \\
   & \quad
       - a^2 \left( \varphi \delta_{ij}
       + \gamma_{,ij} \right) \dot \xi^{,i} \dot \xi^{,j}
       + a^2 \xi_{,ij} \dot \xi^{,i} \dot \xi^{,j}
       + a^2 \dot \xi_{,ij} \dot \xi^{,i} \xi^{,j} \, ,
   \label{GT-alpha}
   \\
   \hat \delta & = \delta
       - \delta_{,i} \xi^{,i}
       + \delta_{,i} \xi^{,i}{}_{,j} \xi^{,j}
       + {1 \over 2} \delta_{,ij} \xi^{,i} \xi^{,j} \, ,
   \label{GT-delta}
   \\
   \hat \kappa & = \kappa
       - \kappa_{,i} \xi^{,i}
       + \kappa_{,i} \xi^{,i}{}_{,j} \xi^{,j}
       + {1 \over 2} \kappa_{,ij} \xi^{,i} \xi^{,j} 
   \label{GT-kappa}
\end{align}
to third order,
\begin{align}
   \hat \beta_{,i} &
       = \beta_{,i}
       + a \dot \xi_{,i}
       - \beta_{,j} \xi^{,j}{}_{,i}
       - \beta_{,ij} \xi^{,j}
       + 2 a \left( \varphi \delta_{ij}
       + \gamma_{,ij} \right) \dot \xi^{,j}
       - a \dot \xi_{,j} \xi^{,j}{}_{,i}
       - a \xi_{,ij} \dot \xi^{,j}
       - a \dot \xi_{,ij} \xi^{,j} \, ,
   \label{GT-beta} 
   \\
   \left( \nabla_i \nabla_j
       - {1 \over 3} \delta_{ij} \Delta \right) \hat \gamma
       & = \left( \nabla_i \nabla_j
       - {1 \over 3} \delta_{ij} \Delta \right) \left( \gamma
       - \xi \right)
       - 2 \varphi \left( \nabla_i \nabla_j
       - {1 \over 3} \delta_{ij} \Delta \right) \xi
       - 2 \gamma_{,k(i} \xi^{,k}{}_{,j)}
       - \gamma_{,ijk} \xi^{,k}
   \nonumber \\
   & \quad
       + {3 \over 2} \xi_{,ik} \xi^{,k}{}_{,j}
       + \xi_{,ijk} \xi^{,k}
       + {1 \over 3} \delta_{ij}
       \left( 2 \gamma^{,k\ell} \xi_{,k\ell}
       + \Delta \gamma_{,k} \xi^{,k}
       - {3 \over 2} \xi^{,k\ell} \xi_{,k\ell}
       - \Delta \xi_{,k} \xi^{,k} \right)
   \label{GT-gamma}
\end{align}
to second order, and 
\begin{equation}
   \hat \varphi = \varphi
\end{equation}
to linear order.

\subsection{Comoving gauge}

The CG condition imposes $\gamma = 0$ in all coordinates.
Eq.~(\ref{GT-gamma}) leads to $\xi = 0$ to second order; apparently, this is true to third order as well, although we do not need it.
Therefore, the CG fixes the gauge degrees of freedom completely.

\subsection{Synchronous gauge}

Whereas, in the SG we impose $\alpha = \beta = 0$ in all coordinates, and Eq.\ (\ref{GT-beta}) leads to
\begin{equation}
   \xi = \xi ({\bf x})
\end{equation}
to second order; again, this is true to third order as
well.
In this way, there remains a gauge degree to second
order even after imposing the SG condition. This is
the notorious remnant gauge mode in the SG. Thus, even
after imposing the SG, we still have coordinate
(gauge) dependent behavior among the synchronous coordinates. 
All the terms involving $\xi$ in 
Eqs.\ (\ref{GT-delta}), (\ref{GT-kappa}) and (\ref{GT-gamma})
are the
remnant gauge modes even after fixing the SG
condition. As we have $\xi = \xi ({\bf x})$, although the values of
$\delta$ and $\kappa$ could still depend on coordinates, the temporal
dependences are the same. That is, the gauge modes behave as 
\begin{equation}
    \delta_G \propto \delta \quad \text{and} \quad
        \kappa_G \propto \kappa \, .
\end{equation}
This explains why we are still able to have second order
differential equations (to third order perturbations) for either 
$\delta$ or $\kappa$ despite
the remaining gauge modes. Apparently, the gauge dependence of $\gamma$,
however, is more complicated especially to second order.
Considering the conserved behavior of $\varphi$ to linear order [see Eq.\ (\ref{SG-sol-linear}) and above],
the gauge mode in $\gamma$ behaves as $\gamma_G \propto
{\rm constant}$ to linear order, and $\gamma_G \propto {\rm
constant}$ and $\gamma$ to second order.

We can show that Eqs.\ (\ref{delta-eq-SG})-(\ref{chi-SG}) remain
valid under the above gauge transformation between any two
SG systems.

\subsection{From CG to SG}

We consider a gauge transformation from the CG to the SG, with the latter being denoted by a hat notation. Thus, we impose $\gamma \equiv 0$ and $\hat \alpha = \hat \beta = 0$ in Eqs.\ (\ref{GT-alpha})-(\ref{GT-gamma}). We have
\begin{align}
   \hat \delta & = \delta
       - \delta_{,i} \xi^{,i}
       + \delta_{,i} \xi^{,i}{}_{,j} \xi^{,j}
       + {1 \over 2} \delta_{,ij} \xi^{,i} \xi^{,j} \, ,
   \\
   \hat \kappa & = \kappa
       - \kappa_{,i} \xi^{,i}
       + \kappa_{,i} \xi^{,i}{}_{,j} \xi^{,j}
       + {1 \over 2} \kappa_{,ij} \xi^{,i} \xi^{,j}
\end{align}
to third order, and 
\begin{align}
   0 & = \beta_{,i}
       + a \dot \xi_{,i}
       - 2 \varphi \beta_{,i}
       + \xi_{,ij} \beta^{,j} \, ,
   \label{CG-SG-beta} 
   \\
   \left( \nabla_i \nabla_j
       - {1 \over 3} \delta_{ij} \Delta \right)
       \left( \hat \gamma + \xi \right)
       & =
       - 2 \varphi \left( \nabla_i \nabla_j
       - {1 \over 3} \delta_{ij} \Delta \right) \xi
       + {3 \over 2} \xi_{,ik} \xi^{,k}{}_{,j}
       + \xi_{,ijk} \xi^{,k}
       - {1 \over 3} \delta_{ij}
       \left( {3 \over 2} \xi^{,k\ell} \xi_{,k\ell}
       + \Delta \xi_{,k} \xi^{,k} \right)
   \label{CG-SG-gamma1}
\end{align}
to second order. 
Here, $\xi \equiv \xi_{\rm CG \to SG}$.

Eq. (\ref{CG-SG-beta}) determines $\xi$ to second order.
However, we notice that even to linear order we have 
from Eq.~\eqref{CG-SG-beta}
\begin{equation}
   \xi = - \int^t {\beta \over a} dt \,,
\end{equation}
and the lower bound of the integration gives rise to the gauge mode $\xi_G$, the remnant in the SG. We have $\xi_G = \xi_G ({\bf x})$. Now, to second order, Eq.\ (\ref{CG-SG-beta}) gives 
\begin{equation}
   \xi = - \int^t {1 \over a} \left[
       \left( 1 - 2 \varphi \right) \beta_{,i}
       + \xi_{,ij} \beta^{,j} \right] dt \, ,
\end{equation}
and the lower bound again gives rise to the remnant gauge mode even to second order.
Thus, we have
\begin{equation}
   \xi_G = \xi_G ({\bf x})
\end{equation}
to second order.
Considering the conserved behavior of $\varphi$ to linear order [see Eq.\ (\ref{SG-sol-linear}) and above], the behaviors of the gauge modes of the SG variables are the following:
\begin{equation}
   \hat \delta_G \propto \delta \, , \quad
       \hat \kappa_G \propto \kappa \, , \quad
       \hat \gamma_G \propto {\rm constant} \, .
\end{equation}
Thus, behaviors of the gauge modes of $\delta$ and $\kappa$ in the SG are the same as the physical modes in the same gauge. This explains why we still have second order differential equations for $\delta$ and $\kappa$ in the SG despite the presence of the remnant gauge modes.

We can derive the equations in the CG from those in the SG using the above gauge transformation.

\subsection{From SG to CG}

We consider a gauge transformation from the SG to the CG, with the latter being denoted by a hat notation. Thus, we impose $\alpha = \beta = 0$ and $\hat \gamma = 0$ in Eqs.\ (\ref{GT-alpha})-(\ref{GT-gamma}). We have 
\begin{align}
   \hat \alpha & = - {1 \over 2} a^2 \dot \xi^{,i} \xi_{,i}
       - {1 \over a^2} \varphi \chi^{,i} \chi_{,i}
       + {1 \over a^2} \chi_{,ij} \chi^{,i} \gamma^{,j},
   \\
   \hat \delta & = \delta
       - \delta_{,i} \xi^{,i}
       + \delta_{,i} \gamma^{,i}{}_{,j} \gamma^{,j}
       + {1 \over 2} \delta_{,ij} \gamma^{,i} \gamma^{,j} \, ,
   \\
   \hat \kappa & = \kappa
       - \kappa_{,i} \xi^{,i}
       + \kappa_{,i} \gamma^{,i}{}_{,j} \gamma^{,j}
       + {1 \over 2} \kappa_{,ij} \gamma^{,i} \gamma^{,j}
\end{align} 
to third order, and
\begin{align}
   \hat \beta_{,i} &
       = a \dot \xi_{,i}
       + {2 \over a} \varphi \chi_{,i}
       - {1 \over a} \chi_{,ij} \gamma^{,j} \, ,
   \\
   0 & = \left( \nabla_i \nabla_j
       - {1 \over 3} \delta_{ij} \Delta \right)
       \left( \gamma - \xi \right)
       - 2 \varphi \left( \nabla_i \nabla_j
       - {1 \over 3} \delta_{ij} \Delta \right) \gamma
       - {1 \over 2} \gamma_{,ik} \gamma^{,k}{}_{,j}
       + {1 \over 6} \delta_{ij}
       \gamma^{,k\ell} \gamma_{,k\ell}
   \label{SG-CG-gamma} 
\end{align}
to second order.
Here, $\xi \equiv \xi_{\rm SG \to CG}$.
Eq.~(\ref{SG-CG-gamma}) determines $\xi$, and this is already
used in the other relations: to linear order we have $\xi =
\gamma$.

We can show that the gauge modes in the SG disappear as we go to the CG.
We can derive the equations in the SG from those in the CG using the above gauge transformation.

\section{Mode analysis in the synchronous gauge}
                                        \label{sec:mode-solutions}

Here we solve the fluid equations in the SG, 
Eqs.\ (\ref{dot-delta-eq-k})-(\ref{X-eq-k}), at each order to find out the 
SG kernels for density $F_{n}$
and velocity $G_n$. Note that, the final expression for the
kernels must be symmetrized over the arguments.

\subsection{Linear order solutions}

In linear order, the equations become
\begin{align}
\dot{\delta}(\bfk,t)
+
\theta(\bfk,t)
& = 0 \, ,
\\
\dot\theta(\bfk,t)
+
2H\theta(\bfk,t)
+
4\pi G\rho \delta(\bfk,t)
& = 0 \,.
\end{align}
These equations in linear order are the same as those of the
Newtonian linear perturbation theory. Therefore, we simply
write down the solutions in the standard way:
\begin{align}
\delta(\bfk,t) & = D(t) \delta_1(\bfk) \,, 
\\
\theta(\bfk,t) & = -H(t)f(t)D(t) \delta_1(\bfk) \, ,
\end{align}
where $D(t)$ is the growth factor, and 
\begin{equation}
f \equiv \frac{d\log D}{d\log a}
\end{equation}
is the logarithmic derivative of the growth factor so that $\dot{D}=HfD$.
Plugging the solutions back to the linearized Euler equation,
we find the following identity:
\begin{equation}
\frac{d(Hf)}{dt}
=
-H^2f^2 - 2H^2f + 4\pi G \rho \, .
\end{equation}
In the EdS universe,
$D(t)=a(t)$ and $f\equiv 1$.
Also, later in higher order, we use the potential perturbation,
whose linear solution is given by
\begin{equation}
\varphi(\bfk,t)
=
\frac{5}{2}\frac{k_H^2}{k^2}\delta(\bfk,t)
=
\frac{5}{2}\frac{k_H^2}{k^2}D(t)\delta_1(\bfk) \, .
\end{equation}

\subsection{Higher order solutions}
We shall find the higher order solutions in terms of the linear
density contrast $\delta_1(\bfk)$.
The usual ansatz for finding such a solution
is writing down the higher order moments as
\begin{align}
\delta(\bfk,t)
& =
\sum_{n=1}^{\infty}
D^n
\int
\frac{d^3q_1}{(2\pi)^3}
\cdots
\int
\frac{d^3q_{n-1}}{(2\pi)^3}
\int d^3q_n
\delta^{(3)}\left(\bfk-\sum_{i=1}^{n}\bfq_i\right)
F_n(\bfq_1,\cdots,\bfq_n,t)
\delta_1(\bfq_1)\cdots
\delta_1(\bfq_n) \, ,
\\
\theta(\bfk,t)
& =
-Hf
\sum_{n=1}^{\infty}
D^n
\int
\frac{d^3q_1}{(2\pi)^3}
\cdots
\int
\frac{d^3q_{n-1}}{(2\pi)^3}
\int d^3q_n
\delta^{(3)}\left(\bfk-\sum_{i=1}^{n}\bfq_i\right)
G_n(\bfq_1,\cdots,\bfq_n,t)
\delta_1(\bfq_1)\cdots
\delta_1(\bfq_n) \, .
\end{align}
We also define $\eta(\bfk,t)$ by
\begin{equation}
\theta(\bfk,t) = -Hf \eta(\bfk,t) \, .
\end{equation}

\subsection{Second order solutions}

The second order equations are
\begin{align}
\dot{\delta}(\bfk,t)
+
\theta(\bfk,t)
& =
-\int \frac{d^3q_1}{(2\pi)^3}
\int d^3q_2
\delta^{(3)}(\bfk-\bfq_{12})
\delta(\bfq_1,t)
\theta(\bfq_2,t) \,,
\\
\dot\theta(\bfk,t)
+
2H\theta(\bfk,t)
+
4\pi G\rho \delta(\bfk,t)
& =
-
\int \frac{d^3q_1}{(2\pi)^3}
\int d^3q_2
\delta^{(3)}(\bfk-\bfq_{12})
\theta(\bfq_1,t)
\theta(\bfq_2,t)\frac{(\bfq_1\cdot\bfq_2)^2}{q_1^2q_2^2} \,.
\end{align}
Plugging the perturbation theory ansatz into the equations and
selecting only second order terms lead to the equations for
$F_2$ and $G_2$.
Note that at this point, we do not symmetrize the kernels yet.
From the second order continuity and Euler equations, we find 
the equations for the kernels as 
\begin{align}
\frac{dF_2}{dt}
+ 2 Hf F_2
- Hf G_2
& =
Hf \, ,
\\
-Hf\frac{dG_2}{dt}
-\frac{\rho}{2}G_2
-H^2f^2G_2
+4\pi G \rho F_2
& =
-H^2f^2\frac{(\bfq_1\cdot\bfq_2)^2}{q_1^2q_2^2} \, .
\end{align}
Finally, by using the Friedman equation
$3H^2 = 8\pi G \rho$
and $f=1$ in the EdS universe,
we find solutions as
\begin{align}
F_2(\bfq_1,\bfq_2)&=
\frac{1}{7}
\left[
5 + 2 \frac{(\bfq_1\cdot\bfq_2)^2}{q_1^2q_2^2}
\right] \, ,
\\
G_2(\bfq_1,\bfq_2) &=
\frac{1}{7}
\left[
3 + 4 \frac{(\bfq_1\cdot\bfq_2)^2}{q_1^2q_2^2}
\right] \, .
\end{align}
Note that these kernels are already symmetric.

\subsection{Third order solutions}

The third order equations are
\begin{align}
\dot{\delta}(\bfk,t)
+
\theta(\bfk,t)
& =
-
\int \frac{d^3q_1}{(2\pi)^3}
\int d^3q_2
\delta^{(3)}(\bfk-\bfq_{12})
\delta(\bfq_1,t)
\theta(\bfq_2,t) \, ,
\\
\dot\theta(\bfk,t)
+
2H\theta(\bfk,t)
+
4\pi G\rho \delta(\bfk,t)
& =
-
\int \frac{d^3q_1}{(2\pi)^3}
\int d^3q_2
\delta^{(3)}(\bfk-\bfq_{12})
\theta(\bfq_1,t)
\theta(\bfq_2,t)\frac{(\bfq_1\cdot\bfq_2)^2}{q_1^2q_2^2}
\nonumber
\\
& \quad +4
\int \frac{d^3q_1}{(2\pi)^3}
\int \frac{d^3q_2}{(2\pi)^3}
\int d^3q_3
\delta^{(3)}(\bfk-\bfq_{123})
\vs
& \qquad \times\left\{
\theta(\bfq_1,t)
\theta(\bfq_2,t)
\varphi(\bfq_3,t)
\left[
\frac{(\bfq_1\cdot\bfq_2)^2}{q_1^2q_2^2}
- {\frac13}
\right]
\right. 
\nonumber\\
& \qquad \qquad {-}\left.
\theta(\bfq_1,t)
\theta(\bfq_2,t)
\delta(\bfq_3,t)
\left[
\frac{(\bfq_1\cdot\bfq_2)(\bfq_2\cdot\bfq_3)(\bfq_3\cdot\bfq_1)}{q_1^2q_2^2q_3^2}
-
\frac{1}{3}\frac{(\bfq_1\cdot\bfq_3)^2}{q_1^2q_3^2}
\right]
\right\}
\nonumber\\
&\quad +
2\int \frac{d^3q_1}{(2\pi)^3}
\int d^3q_2
\delta^{(3)}(\bfk-\bfq_{12})
\theta(\bfq_1,t)
X(\bfq_2,t)
\left[
\frac{1}{3}
-
\frac{(\bfq_1\cdot\bfq_2)^2}{q_1^2q_2^2}
\right] \, ,
\end{align}
where
\begin{align}
X(\bfk,t)
& =
\int \frac{d^3q_1}{(2\pi)^3}
\int d^3q_2
\delta^{(3)}(\bfk-\bfq_{12})
\nonumber
\\
& \quad \times
\left\{
\theta(\bfq_1,t)\varphi(\bfq_2,t)
\left[
2-\frac{3}{2}\frac{\bfq_1\cdot\bfq_2}{q_1^2}
+\frac{3}{2}\frac{\bfq_{12}\cdot\bfq_2}{q_{12}^2}
\left(
-1+\frac{\bfq_1\cdot\bfq_2}{q_1^2}
\right)
\right]
\right.
\vs
&\qquad\quad+
\left.
\theta(\bfq_1,t)\delta(\bfq_2,t)
\left[
-\frac{3}{2}\frac{\bfq_1\cdot\bfq_2}{q_1^2}
-2\frac{(\bfq_1\cdot\bfq_2)^2}{q_1^2q_2^2}
+
\frac{3}{2}\frac{(\bfq_{12}\cdot\bfq_2)^2(\bfq_1\cdot\bfq_2)}{q_{12}^2q_1^2q_2^2}
\right]
\right\} \, .
\end{align}
Because we only need the right hand side of the Euler equation
up to third order, we first calculate $X(\bfk,t)$ up to second
order by using the linear solutions in EdS universe:
\begin{align}
X(\bfk,t)
=&
-HD^2
\int \frac{d^3q_1}{(2\pi)^3}
\int d^3q_2
\delta^{(3)}(\bfk-\bfq_{12})
\delta_1(\bfq_1)
\delta_1(\bfq_2)
\nonumber
\\
&\times
\left\{
\frac{5}{2}\frac{k_H^2}{q_2^2}
\left[
2-\frac{3}{2}\frac{\bfq_1\cdot\bfq_2}{q_1^2}
+\frac{3}{2}\frac{\bfq_{12}\cdot\bfq_2}{q_{12}^2}
\left(
-1+\frac{\bfq_1\cdot\bfq_2}{q_1^2}
\right)
\right]
+
\left[
-\frac{3}{2}\frac{\bfq_1\cdot\bfq_2}{q_1^2}
-2\frac{(\bfq_1\cdot\bfq_2)^2}{q_1^2q_2^2}
+
\frac{3}{2}\frac{(\bfq_{12}\cdot\bfq_2)^2(\bfq_1\cdot\bfq_2)}{q_{12}^2q_1^2q_2^2}
\right]
\right\} \, .
\end{align}
By using the perturbation theory ansatz, the third order 
continuity and Euler equations are reduced to 
\begin{align}
&
\frac{1}{H}\frac{dF_3}{dt} + 3 F_3 -  G_3
=
F_2(\bfq_1,\bfq_3)
+
G_2(\bfq_2,\bfq_3) \, ,
\\
\nonumber
&\frac{1}{H}\frac{dG_3}{dt} + 2 G_3 - \frac{3}{2}(F_3-G_3)
\\
\nonumber
=&
2\frac{(\bfq_1\cdot\bfq_{23})^2}{q_1^2q_{23}^2}G_2(\bfq_2,\bfq_3)
 {+}
4
\left[
\frac{(\bfq_1\cdot\bfq_2)(\bfq_2\cdot\bfq_3)(\bfq_3\cdot\bfq_1)}{q_1^2q_2^2q_3^2}
-
\frac{1}{3}\frac{(\bfq_1\cdot\bfq_3)^2}{q_1^2q_3^2}
\right]
\nonumber
\\
\nonumber
&+
\left[
3\frac{\bfq_2\cdot\bfq_3}{q_2^2}
+4\frac{(\bfq_2\cdot\bfq_3)^2}{q_2^2q_3^2}
-
3\frac{(\bfq_{23}\cdot\bfq_3)^2(\bfq_2\cdot\bfq_3)}{q_{23}^2q_2^2q_3^2}
\right]
\left[
\frac{1}{3}
-
\frac{(\bfq_1\cdot\bfq_{23})^2}{q_1^2q_{23}^2}
\right]
\\
&-
\left\{
5\frac{k_H^2}{q_3^2}
\left[
2-\frac{3}{2}\frac{\bfq_2\cdot\bfq_3}{q_2^2}
+\frac{3}{2}\frac{\bfq_{23}\cdot\bfq_3}{q_{23}^2}
\left(
-1+\frac{\bfq_2\cdot\bfq_3}{q_2^2}
\right)
\right]
\left[
\frac{1}{3}
-
\frac{(\bfq_1\cdot\bfq_{23})^2}{q_1^2q_{23}^2}
\right]
+
10\frac{k_H^2}{q_3^2}
\left[
\frac{(\bfq_1\cdot\bfq_2)^2}{q_1^2q_2^2}
- {\frac13}
\right]
\right\} \, .
\end{align}
We divide the solutions for the kernels by time-dependent parts
(proportional to $k_H$) and 
time-independent parts (with superscript \textit{tid}).
The time-independent parts of the kernels are
\begin{align}
F_3^{tid}(\bfq_1,\bfq_2,\bfq_3)
& =
\frac{7}{18}
\left[
F_2(\bfq_1,\bfq_3)
+
G_2(\bfq_2,\bfq_3)
\right]
+
\frac{2}{9}\frac{(\bfq_1\cdot\bfq_{23})^2}{q_1^2q_{23}^2}G_2(\bfq_2,\bfq_3),
\nonumber
\\
& \quad
 {+}
\frac{4}{9}
\left[
\frac{(\bfq_1\cdot\bfq_2)(\bfq_2\cdot\bfq_3)(\bfq_3\cdot\bfq_1)}{q_1^2q_2^2q_3^2}
-
\frac{1}{3}\frac{(\bfq_1\cdot\bfq_3)^2}{q_1^2q_3^2}
\right]
\nonumber
\\
& \quad
+\frac{1}{9}
\left[
3\frac{\bfq_2\cdot\bfq_3}{q_2^2}
+4\frac{(\bfq_2\cdot\bfq_3)^2}{q_2^2q_3^2}
-
3\frac{(\bfq_{23}\cdot\bfq_3)^2(\bfq_2\cdot\bfq_3)}{q_{23}^2q_2^2q_3^2}
\right]
\left[
\frac{1}{3}
-
\frac{(\bfq_1\cdot\bfq_{23})^2}{q_1^2q_{23}^2}
\right] \, ,
\\
G_3^{tid}(\bfq_1,\bfq_2,\bfq_3)
& =
\frac{1}{6}
\left[
F_2(\bfq_1,\bfq_3)
+
G_2(\bfq_2,\bfq_3)
\right]
+
\frac{2}{3}\frac{(\bfq_1\cdot\bfq_{23})^2}{q_1^2q_{23}^2}G_2(\bfq_2,\bfq_3)
\nonumber
\\
& \quad
 {+}\frac{4}{3}
\left[
\frac{(\bfq_1\cdot\bfq_2)(\bfq_2\cdot\bfq_3)(\bfq_3\cdot\bfq_1)}{q_1^2q_2^2q_3^2}
-
\frac{1}{3}\frac{(\bfq_1\cdot\bfq_3)^2}{q_1^2q_3^2}
\right]
\nonumber
\\
& \quad
+\frac{1}{3}
\left[
3\frac{\bfq_2\cdot\bfq_3}{q_2^2}
+4\frac{(\bfq_2\cdot\bfq_3)^2}{q_2^2q_3^2}
-
3\frac{(\bfq_{23}\cdot\bfq_3)^2(\bfq_2\cdot\bfq_3)}{q_{23}^2q_2^2q_3^2}
\right]
\left[
\frac{1}{3}
-
\frac{(\bfq_1\cdot\bfq_{23})^2}{q_1^2q_{23}^2}
\right] \, .
\end{align}
The time-independent kernels above can be further divided by two parts: 
(purely) Newtonian terms and relativistic terms.
The first three terms are pure Newtonian as we can identify them as
the third order kernels of the Lagrangian fluid equations (see Appendix \ref{sec:Eulerian-modification}), and the rest of terms come from the non-linear coupling due to 
the non-linear nature of the Einstein's gravity. 
Note that although general relativistic, 
those terms do not include the gravitational potential $\varphi$ and thus are 
not multiplied by $k_H^2$:
\begin{align}
F_3^{(N)}(\bfq_1,\bfq_2,\bfq_3)
& =
\frac{7}{18}
\left[
F_2(\bfq_1,\bfq_3)
+
G_2(\bfq_2,\bfq_3)
\right]
+
\frac{2}{9}\frac{(\bfq_1\cdot\bfq_{23})^2}{q_1^2q_{23}^2}
G_2(\bfq_2,\bfq_3) \, ,
\\
F_{3,\mathrm{no}~\varphi}^{(E)}(\bfq_1,\bfq_2,\bfq_3)
& =
\frac{4}{9}
\left[
\frac{(\bfq_1\cdot\bfq_2)(\bfq_2\cdot\bfq_3)(\bfq_3\cdot\bfq_1)}{q_1^2q_2^2q_3^2}
-
\frac{1}{3}\frac{(\bfq_1\cdot\bfq_3)^2}{q_1^2q_3^2}
\right]
\vs
& \quad +\frac{1}{9}
\left[
3\frac{\bfq_2\cdot\bfq_3}{q_2^2}
+4\frac{(\bfq_2\cdot\bfq_3)^2}{q_2^2q_3^2}
-
3\frac{(\bfq_{23}\cdot\bfq_3)^2(\bfq_2\cdot\bfq_3)}{q_{23}^2q_2^2q_3^2}
\right]
\left[
\frac{1}{3}
-
\frac{(\bfq_1\cdot\bfq_{23})^2}{q_1^2q_{23}^2}
\right] \, ,
\\
G_3^{(N)}(\bfq_1,\bfq_2,\bfq_3)
& =
\frac{1}{6}
\left[
F_2(\bfq_1,\bfq_3)
+
G_2(\bfq_2,\bfq_3)
\right]
+
\frac{2}{3}\frac{(\bfq_1\cdot\bfq_{23})^2}{q_1^2q_{23}^2}G_2(\bfq_2,\bfq_3) \, ,
\\
G_{3,\mathrm{no}~\varphi}^{(E)}(\bfq_1,\bfq_2,\bfq_3)
& = \frac{4}{3}
\left[
\frac{(\bfq_1\cdot\bfq_2)(\bfq_2\cdot\bfq_3)(\bfq_3\cdot\bfq_1)}{q_1^2q_2^2q_3^2}
-
\frac{1}{3}\frac{(\bfq_1\cdot\bfq_3)^2}{q_1^2q_3^2}
\right]
\nonumber
\\
& \quad +\frac{1}{3}
\left[
3\frac{\bfq_2\cdot\bfq_3}{q_2^2}
+4\frac{(\bfq_2\cdot\bfq_3)^2}{q_2^2q_3^2}
-
3\frac{(\bfq_{23}\cdot\bfq_3)^2(\bfq_2\cdot\bfq_3)}{q_{23}^2q_2^2q_3^2}
\right]
\left[
\frac{1}{3}
-
\frac{(\bfq_1\cdot\bfq_{23})^2}{q_1^2q_{23}^2}
\right] \, .
\end{align}

The time-dependent parts of the third order kernels that include $\varphi$ are the solutions of 
following differential equations:
\begin{align}
\frac{1}{H}\frac{dF_{3,\varphi}^{(E)}}{dt}
+
3 F_{3,\varphi}^{(E)} - G_{3,\varphi}^{(E)}
& = 0 \, ,
\\
-\frac{1}{H}\frac{dG_{3,\varphi}^{(E)}}{dt}
+
\frac{3}{2}F_{3,\varphi}^{(E)} - \frac{7}{2}G_{3,\varphi}^{(E)}
& =
5\frac{k_H^2}{q_3^2}
\left[
2-\frac{3}{2}\frac{\bfq_2\cdot\bfq_3}{q_2^2}
+\frac{3}{2}\frac{\bfq_{23}\cdot\bfq_3}{q_{23}^2}
\left(
-1+\frac{\bfq_2\cdot\bfq_3}{q_2^2}
\right)
\right]
\left[
\frac{1}{3}
-
\frac{(\bfq_1\cdot\bfq_{23})^2}{q_1^2q_{23}^2}
\right]
\vs
&\quad+
10\frac{k_H^2}{q_3^2}
\left[
\frac{(\bfq_1\cdot\bfq_2)^2}{q_1^2q_2^2}
- {\frac13}
\right] \,,
\end{align}
and we find
\begin{equation}
F_{3,\varphi}^{(E)}(\bfq_1,\bfq_2,\bfq_3)
=
-\frac{10}{7}
\frac{k_H^2}{q_3^2}
\left\{
\left[
2-\frac{3}{2}\frac{\bfq_2\cdot\bfq_3}{q_2^2}
+\frac{3}{2}\frac{\bfq_{23}\cdot\bfq_3}{q_{23}^2}
\left(
-1+\frac{\bfq_2\cdot\bfq_3}{q_2^2}
\right)
\right]
\left[
\frac{1}{3}
-
\frac{(\bfq_1\cdot\bfq_{23})^2}{q_1^2q_{23}^2}
\right]
+
2
\left[
\frac{(\bfq_1\cdot\bfq_2)^2}{q_1^2q_2^2}
- {\frac13}
\right]
\right\} \, ,
\end{equation}
and $G_{3,\varphi}^{(E)} = 2 F_{3,\varphi}^{(E)}$.

\section{Newtonian non-linear power spectrum in the Eulerian and Lagrangian frames}
                                        \label{sec:Eulerian-modification}

In the main text, we have shown that to second order in perturbation, the fluid
equations in the SG coincide with the Lagrangian view of the Newtonian
fluid equations. In this section, we calculate the perturbative kernel 
solutions to third order for the Lagrangian fluid equations. 
The kernels we find here are identified as the Newtonian terms of the 
relativistic solutions we find in Appendix \ref{sec:mode-solutions}.

Let us start from the Eulerian fluid equations:
\begin{align}
&\frac{d\delta}{dt} +\frac1a\nabla\cdot\left[(1+\delta)\bfu\right]  = 0 \, ,
\\
&\frac{d\bfu}{dt} +
\frac1a
\left(\bfu\cdot\nabla\right)\bfu
 =
-H\bfu - \frac1a\nabla{\phi} \, ,
\\
&\Delta\phi
 =
4\pi G  \rho a^2 \delta \, .
\end{align}
We combine the Euler equation and Poisson equation to find 
\begin{equation}
\frac{d\theta}{dt}
+
\frac{1}{a^2}
\nabla\cdot
\left[
\left(\bfu\cdot\nabla\right)\bfu\right]
=
-2 H\theta -
4\pi G  \rho \delta \, ,
\end{equation}
with $\theta \equiv a^{-1}\nabla\cdot\bfu$.
From here, we obtain the Lagrangian fluid equations by changing the
time derivative to the convective derivative:
\begin{equation}
\frac{d}{dt} \to
\frac{D}{Dt} \equiv
\frac{d}{dt} + \frac1a\bfu\cdot\nabla \, ,
\end{equation}
then the fluid equations now become
\begin{align}
&\frac{D\delta}{Dt} + \left(1+\delta\right)\frac1a\nabla\cdot\bfu  = 0 \, ,
\\
&\frac{D\theta}{Dt}
+
\frac{1}{a^2} 
\partial_i u^j
\partial_j u^i
 =
- 2 H\theta -
4\pi G \rho \delta \, .
\end{align}

In the Fourier space, the Eulerian fluid equations become
\begin{align}
\label{eq:Econt}
\frac{d\delta(\bfk,t)}{dt}
+
\theta(\bfk,t)
& =
-
\int
\frac{d^3q_1}{(2\pi)^3}
\int d^3 q_2
\frac{\bfk\cdot\bfq_2}{q_2^2}
\delta(\bfq_1,t)\theta(\bfq_2,t)
\delta^{(3)}(\bfk-\bfq_{12}) \, ,
\\
\label{eq:Eeuler}
\frac{d\theta(\bfk,t)}{dt}
+
2H\theta(\bfk,t)
+
\frac{3}{2}H^2\Omega_m\delta(\bfk,t)
&=
-
\int
\frac{d^3q_1}{(2\pi)^3}
\int d^3 q_2
\frac{k^2(\bfq_1\cdot\bfq_2)}{2q_1^2q_2^2}
\theta(\bfq_1,t)\theta(\bfq_2,t)
\delta^{(3)}(\bfk-\bfq_{12}) \, ,
\end{align}
whereas the Lagrangian fluid equations become
\begin{align}
\label{eq:Lcont}
\frac{D\delta(\bfk,t)}{Dt}
+
\theta(\bfk,t)
& =
-
\int
\frac{d^3q_1}{(2\pi)^3}
\int d^3 q_2
\delta(\bfq_1,t)\theta(\bfq_2,t)
\delta^{(3)}(\bfk-\bfq_{12}) \, ,
\\
\label{eq:Leuler}
\frac{D\theta(\bfk)}{Dt}
+
2H\theta(\bfk,t)
+
\frac{3}{2}H^2\Omega_m\delta(\bfk,t)
& =
-
\int
\frac{d^3q_1}{(2\pi)^3}
\int d^3 q_2
\frac{(\bfq_1\cdot\bfq_2)^2}{q_1^2q_2^2}
\theta(\bfq_1,t)\theta(\bfq_2,t)
\delta^{(3)}(\bfk-\bfq_{12}) \, .
\end{align}
Note that the Eulerian and Lagrangian
fluid equations coincide exactly with the corresponding equations truncated
at second order, respectively, in the CG and the SG.

\subsection{Solutions}
\begin{figure}[t!]
\centering
\includegraphics*[width=0.49\textwidth]{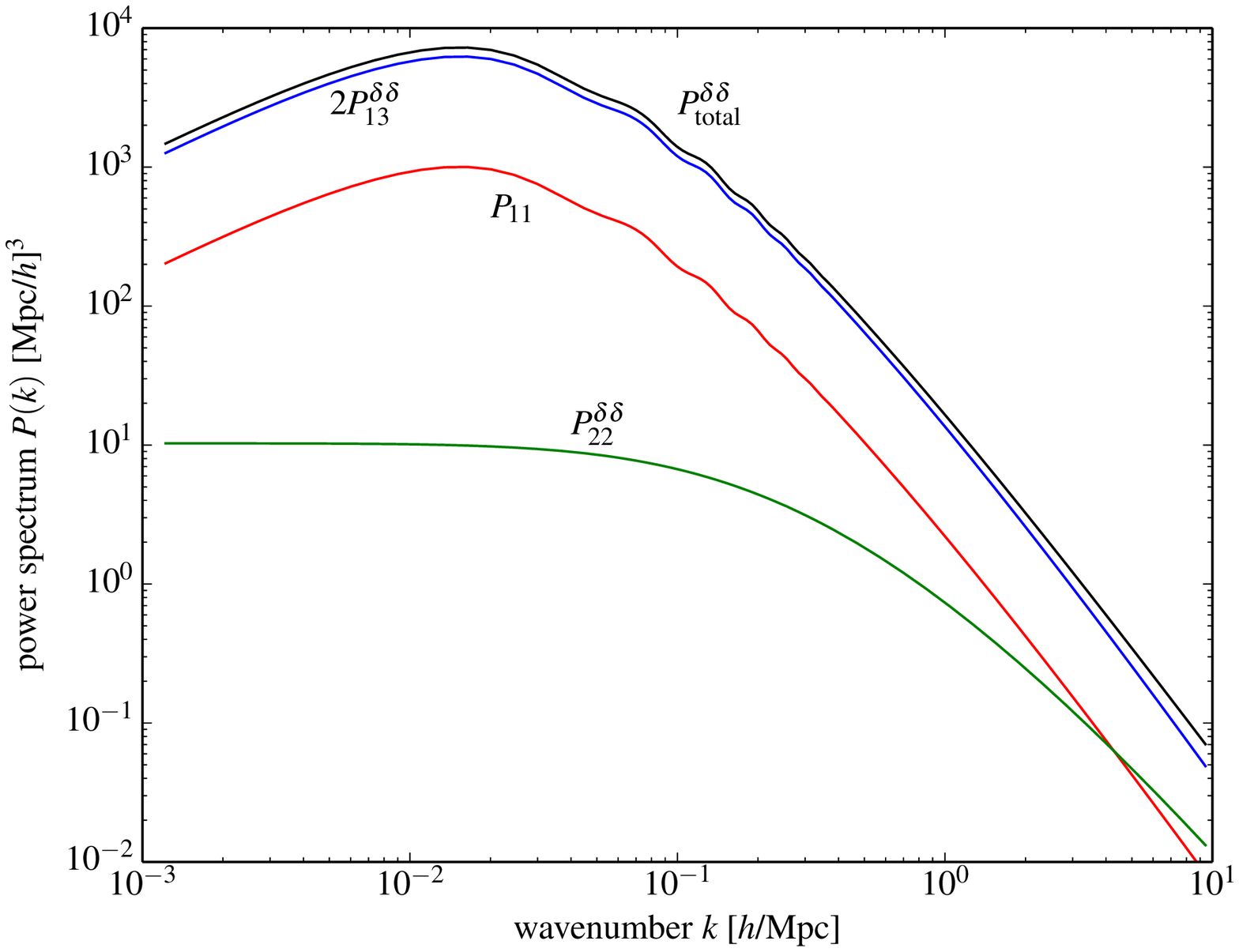}
\includegraphics*[width=0.49\textwidth]{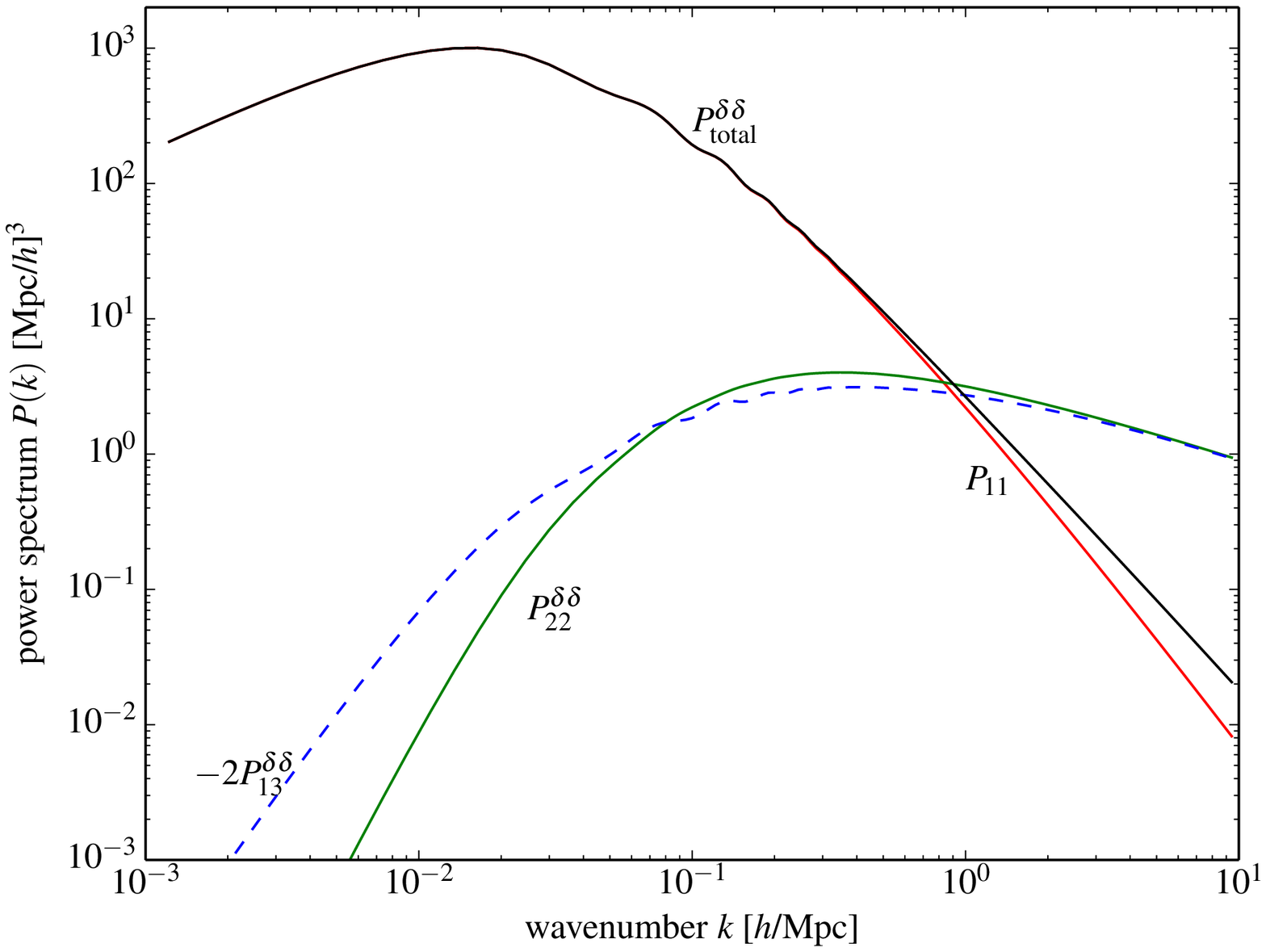}
\includegraphics*[width=0.49\textwidth]{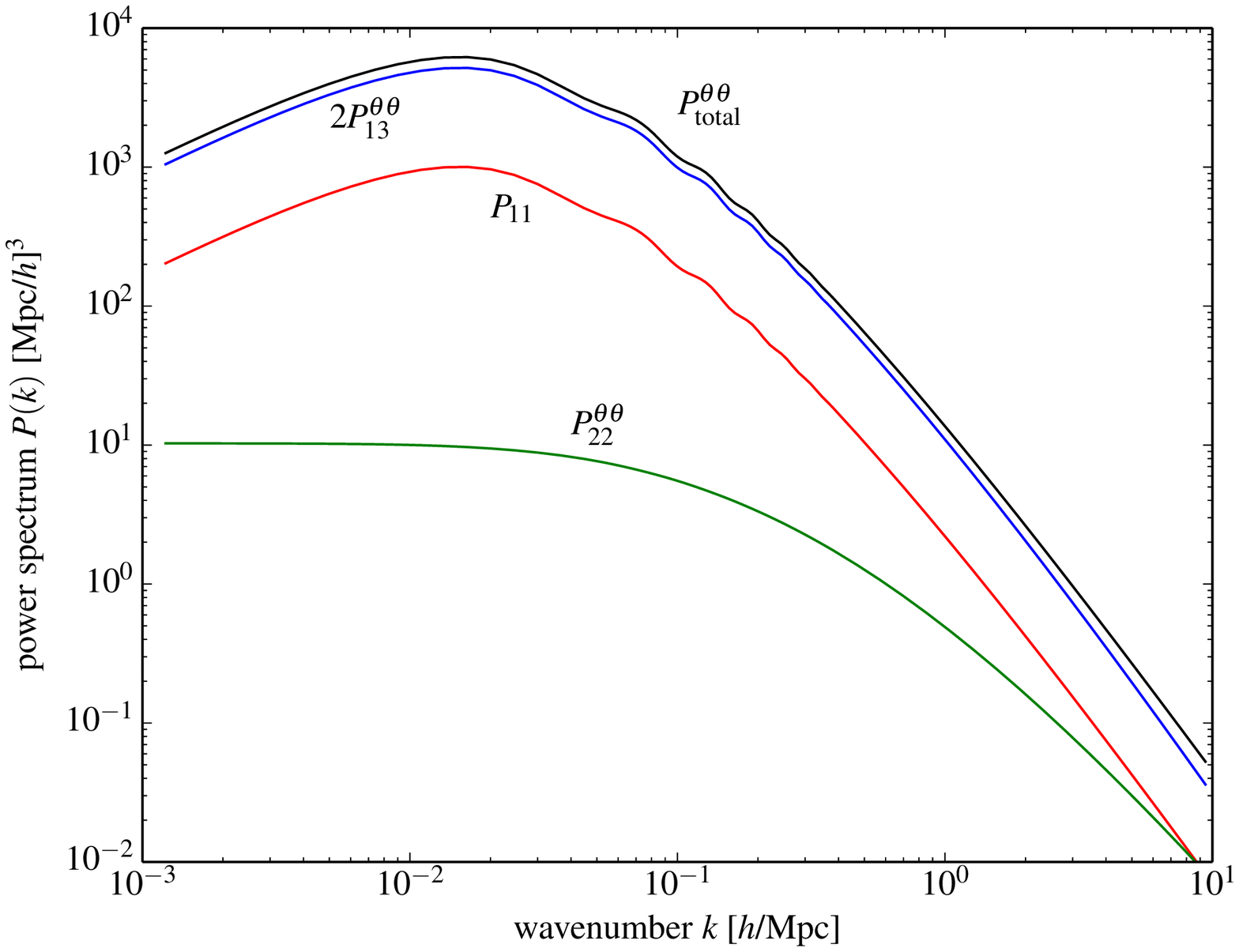}
\includegraphics*[width=0.49\textwidth]{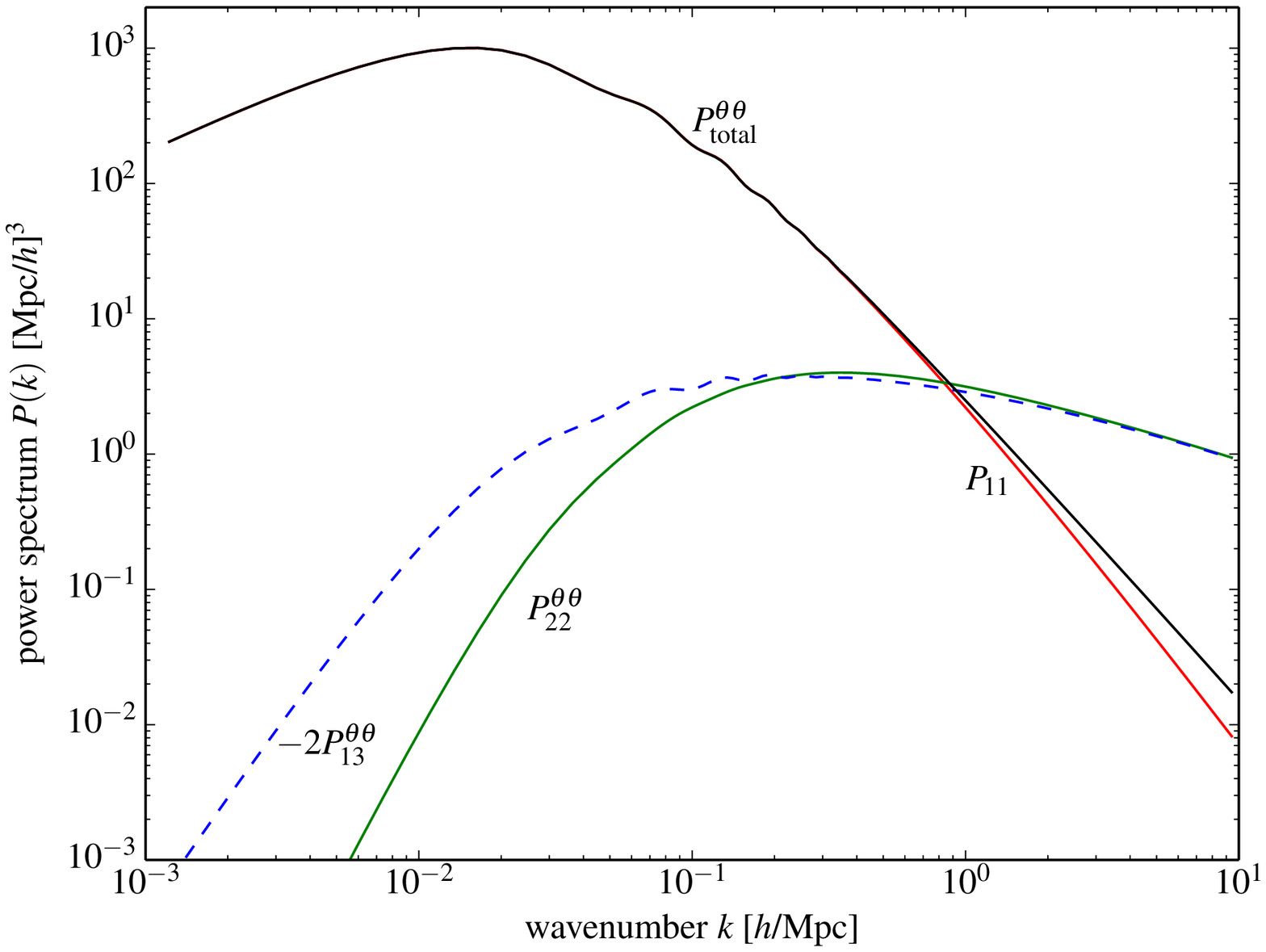}
\caption{
Density (\textit{top panels}) 
and velocity (\textit{bottom panels}) power spectra of the Newtonian 
perturbations calculated in the Lagrangian (\textit{left panels}) and 
Eulerian (\textit{right panels}) frames in the EdS universe at $z=6$. 
The Lagrangian and Eulerian power spectra shown here are identical to the 
\textit{Newtonian} power spectra shown in the main text,  Figures~\ref{fig:pkSC}
and \ref{fig:pkSC_LtoE} respectively.
}
\label{fig:pk_Newtonian}
\end{figure}

In linear order,
the continuity [Eqs.~\eqref{eq:Econt} and \eqref{eq:Lcont}]
and the Euler [Eqs.~\eqref{eq:Eeuler} and \eqref{eq:Leuler}]
equations coincide, and the difference appears from
second order.
The second order solution kernels
$F_{2,E}$ and $G_{2,E}$
for density and velocity gradient in the Eulerian fluid equations
satisfy
\begin{align}
2F_{2,E} - G_{2,E} & =
\frac{\bfk\cdot\bfq_2}{q_2^2} \, ,
\\
\frac{5}{2}G_{2,E} - \frac{3}{2}F_{2,E} & =
\frac{k^2(\bfq_1\cdot\bfq_2)}{2q_1^2q_2^2} \, ,
\end{align}
while
$F_{2,L}$ and $G_{2,L}$
in the Lagrangian fluid equations satisfy
\begin{align}
2F_{2,L} - G_{2,L} & = 1 \, ,
\\
\frac{5}{2}G_{2,L} - \frac{3}{2}F_{2,L} & =
\frac{(\bfq_1\cdot\bfq_2)^2}{q_1^2q_2^2} \, .
\end{align}
For the third order kernels, we have
\begin{align}
3F_{3,E} - G_{3,E} & =
\frac{\bfk\cdot\bfq_2}{q_2^2}
F_{2,E}(\bfq_1,\bfq_3)
+
\frac{\bfk\cdot\bfq_{23}}{q_{23}^2}
G_{2,E}(\bfq_2,\bfq_3) \, ,
\\
\frac{3}{2}F_{3,E} - \frac{7}{2}G_{3,E} & =
-\frac{k^2(\bfq_1\cdot\bfq_{23})}{q_1^2q_{23}^2}
G_{2,E}(\bfq_2,\bfq_3)
\end{align}
for the Eulerian fluid equations, and
\begin{align}
3F_{3,L} - G_{3,L} & =
F_{2,L}(\bfq_1,\bfq_3)
+
G_{2,L}(\bfq_2,\bfq_3),
\\
\frac{3}{2}F_{3,L} - \frac{7}{2}G_{3,L} & =
-\frac{2(\bfq_1\cdot\bfq_{23})^2}{q_1^2q_{23}^2}
G_{2,L}(\bfq_2,\bfq_3)
\end{align}
for the Lagrangian fluid equations.

Solving the equations above, we find the second and third order
solutions as
\begin{align}
F_{2,E}(\bfq_1,\bfq_2)
& =
\frac{1}{7}
\left[
\frac{k^2(\bfq_1\cdot\bfq_2)}{q_1^2q_2^2}
+
5\frac{\bfk\cdot\bfq_2}{q_2^2}
\right] \, ,
\\
G_{2,E}(\bfq_1,\bfq_2)
& =
\frac{1}{7}
\left[
2 \frac{k^2(\bfq_1\cdot\bfq_2)}{q_1^2q_2^2}
+
3 \frac{\bfk\cdot\bfq_2}{q_2^2}
\right] \, ,
\\
F_{2,L}(\bfq_1,\bfq_2)
& =
\frac{1}{7}
\left[
2 \frac{(\bfq_1\cdot\bfq_2)^2}{q_1^2q_2^2}
+ 
5
\right] \, ,
\\
G_{2,L}(\bfq_1,\bfq_2)
& =
\frac{1}{7}
\left[
4 \frac{(\bfq_1\cdot\bfq_2)^2}{q_1^2q_2^2}
+
3
\right]\,,
\end{align}
\begin{align}
F_{3,E} &=
\frac{1}{18}
\left[
7
\frac{\bfk\cdot\bfq_2}{q_2^2}F_{2,E}(\bfq_1,\bfq_3)
+
7
\frac{\bfk\cdot\bfq_{23}}{q_{23}^2}G_{2,E}(\bfq_2,\bfq_3)
+
2
\frac{k^2(\bfq_1\cdot\bfq_{23})}{q_1^2q_{23}^2}G_{2,E}(\bfq_2,\bfq_3)
\right] \, ,
\\
G_{3,E} &=
\frac{1}{6}
\left[
\frac{\bfk\cdot\bfq_2}{q_2^2}F_{2,E}(\bfq_1,\bfq_3)
+
\frac{\bfk\cdot\bfq_{23}}{q_{23}^2}G_{2,E}(\bfq_2,\bfq_3)
+2
\frac{k^2(\bfq_1\cdot\bfq_{23})}{q_1^2q_{23}^2}G_{2,E}(\bfq_2,\bfq_3)
\right] \, ,
\\
F_{3,L} &=
\frac{1}{18}
\left[
7
F_{2,L}(\bfq_1,\bfq_3)
+
7
G_{2,L}(\bfq_2,\bfq_3)
+
4
\frac{(\bfq_1\cdot\bfq_{23})^2}{q_1^2q_{23}^2}G_{2,L}(\bfq_2,\bfq_3)
\right] \, ,
\\
G_{3,L} &=
\frac{1}{6}
\left[
F_{2,L}(\bfq_1,\bfq_3)
+
G_{2,L}(\bfq_2,\bfq_3)
+4
\frac{(\bfq_1\cdot\bfq_{23})^2}{q_1^2q_{23}^2}G_{2,L}(\bfq_2,\bfq_3)
\right] \, .
\end{align}

We show the next-to-leading order matter (\textit{top panel}) and velocity
(\textit{bottom panel}) power spectra in Figure~\ref{fig:pk_Newtonian} for both the
Lagrangian (\textit{left panel}) and Eulerian (\textit{right panel}) 
coordinate frames. Note that the Lagrangian and Eulerian power spectra shown 
here are identical to the \textit{Newtonian} power spectra shown in 
Figures~\ref{fig:pkSC} and \ref{fig:pkSC_LtoE} respectively.
That is, the irregular behavior of the non-linear power spectrum 
is already apparent in the Newtonian perturbation theory when using an
inappropriate Lagrangian coordinate frame.

%
%


\end{document}